**Disentangling basal and accrued height-for-age for cross-population comparisons**

Joseph V. Hackman[1], Daniel J. Hruschka[2]

[1]Department of Anthropology, University of Utah, Salt Lake City, UT
[2]School of Human Evolution and Social Change, Arizona State University, Tempe, AZ

Correspondence concerning this article should be addressed to:

Joseph Hackman
DEPARTMENT OF ANTHROPOLOGY
260 S. CENTRAL CAMPUS DRIVE, RM 4625
SALT LAKE CITY, UT 84112
Email: joehackman@gmail.com

**ABSTRACT**

**Objectives:** Current standards for comparing stunting across human populations assume a universal model of child growth. Such comparisons ignore population differences that are independent of deprivation and health outcomes. This paper partitions variation in height-for-age that is specifically associated with deprivation and health outcomes to provide a basis for cross-population comparisons.

**Materials & Methods:** Using a multi-level model with a sigmoid relationship of resources and growth, we partition variation in height-for-age z-scores (HAZ) from 1,522,564 children across 70 countries into two components: 1) "accrued HAZ" shaped by environmental inputs (e.g., undernutrition, infectious disease, inadequate sanitation, poverty), and 2) a country-specific "basal HAZ" independent of such inputs. We validate these components against population-level infant mortality rates, and assess how these basal differences may affect cross-population comparisons of stunting.

**Results:** Basal HAZ differs reliably across countries (range of 1.5 SD) and is independent of measures of infant mortality. By contrast, accrued HAZ captures stunting as impaired growth due to deprivation and is more closely associated with infant mortality than observed HAZ. Ranking populations by accrued HAZ suggest that populations in West Africa and the Caribbean suffer much greater levels of stunting than suggested by observed HAZ.

**Discussion:** Current universal standards may dramatically underestimate stunting in populations with taller basal HAZ. Relying on observed HAZ rather than accrued HAZ may also lead to inappropriate cross-population comparisons, such as concluding that Haitian children enjoy better conditions for growth than do Indian or Guatemalan children.

Abstract =240, Main Text (5,300), Tables (2), Figures (6), References (64)

## INTRODUCTION

Stunting, or impaired linear growth due to undernutrition and deprivation, has numerous negative consequences for health, cognitive ability, and long-term academic and economic achievement (De Onis, Blössner, & Borghi, 2012; Prendergast & Humphrey, 2014; Victora et al., 2008).  With an estimated 165 million children under the age of 5 currently suffering from stunting (Angood et al., 2016; De Onis et al., 2012; Kim, 2016), monitoring, targeting, and preventing stunting at the individual and population level has become a key global health priority (De Onis et al., 2013; Frongillo, Leroy, & Lapping, 2019; Leroy & Frongillo, 2019; Osgood-Zimmerman et al., 2018; Perumal, Bassani, & Roth, 2018; Roth et al., 2017; Stevens et al., 2012).   The underlying metric for assessing stunting—height-for-age z-scores (HAZ)—is also frequently used in anthropology and the social sciences to compare relative deprivation and healthy growth across individuals and populations (Gaur & Kumar, 2012; Hermanussen, Bogin, & Scheffler, 2018; Schillaci, Sachdev, & Bhargava, 2012; Sterling et al., 2012; Tanner, Leonard, & Reyes-García, 2014; Vercellotti et al., 2014).

It is well-established that environmental factors, such as disease exposure, nutrition, access to healthcare, and other forms of deprivation, are associated with reduced height-for-age (Dewey & Begum, 2011; Hoddinott, Maluccio, Behrman, & Flores, 2008; Schroeder, Martorell, & Rivera, 1995; Waterlow, 1994).   However, relatively stable, non-environmental factors, including genetic variation, can also contribute to variation in height-for-age (Davies, 1988; Goldstein & Tanner, 1980; RONA, 1981; Weedon, Lettre, Freathy, & Lindgren, 2007) (Coffey, Deaton, Dreze, Dean, & Tarozzi, 2013).

Current global standards for comparing height across populations assume that population differences arise primarily from environmental factors (Borghi et al., 2006; De Onis, Onyango, Borghi, Garza, & Yang, 2006; Graitcer & Gentry, 1981; WHO Multicentre growth reference study group, 2006).  According to this model, the growth of healthy, exclusively breastfed children does not vary substantially across populations, and a single set of growth curves is sufficient to describe a universal norm of healthy childhood growth. The World Health Organization's (WHO) Multicentre Growth Reference Study (MGRS) established such a set of curves from sites in six countries—Brazil, Ghana, India, Norway, Oman and the United States. These curves now serve as the WHO's standard target for childhood growth across all countries, with thresholds for classifying stunting based on height-for-age z-scores (HAZ) less than -2SD for moderate stunting and -3SD for severe stunting (De Onis, Garza, Victora, Onyango, & Edward, 2004; De Onis & Yip, 1996; Martorell & Young, 2012; Natale & Rajagopalan, 2014).

Resting on a universal model of human growth, this approach assumes that non-environmental factors contribute to a sufficiently small portion of population differences in height that a single standard can be used to assess healthy growth across all populations. However, empirical studies across a wider range of countries have provided mixed support for this assumption (Buuren & Wouwe, 2008; Christesen, Pedersen, Pournara, Petit, & Júlíusson, 2016; De Wilde, van Dommelen, Van Buuren, & Middelkoop, 2015; Graitcer & Gentry, 1981; Hui, Schooling, & Cowling, 2008; Karra, Subramanian, & Fink, 2017; Panagariya, 2013; Rojroongwasinkul et al., 2016).

For example, across a number of European countries, Christesen et al. (2016) found that WHO growth standards were more likely to misclassify children with growth hormone deficiency than were country-specific standards.

To further assess assumptions underlying universal growth models, we examine a two-component model of height that partitions height-for-age into two components (equation 1). According to this model, the first component of variation in HAZ results from improvements in environmental inputs, such as infectious disease burden, nutrition, access to healthcare, and socioeconomic resources (henceforth, accrued HAZ). The second component of variation in HAZ exists independent of such improvements, and reflects population-specific basal levels (henceforth, basal HAZ).  According to this model, variation in basal HAZ would represent the variation observed between populations when environmental inputs are held constant.  The universal model of growth underlying current WHO standards assumes that variation in basal HAZ is sufficiently small that observed HAZ is a straightforward measure of accrued HAZ.

$$\text{observed HAZ} = \text{accrued HAZ} + \text{basal HAZ} \qquad (1)$$

Evidence from another measure of human growth—weight-for-height—in both children and adults indicates that a basal component independent of environmental inputs can contribute substantially to population variation in human growth (De Wilde, 2013; Hadley & Hruschka, 2014, 2017; D. Hruschka & Hadley, 2016; D. Hruschka, Hadley, Brewis, & Stojanowski, 2015; Pomeroy, Mushrif-Tripathy, Cole, Wells, & Stock, 2019)

Using the two-component model for weight-for-height, one study found that universal cutoffs that ignore basal differences potentially underestimate global rates of overweight by 400-500 million in adults (D. Hruschka, Hadley, & Brewis, 2014) and can dramatically shift rankings of those populations most at risk for wasting in children (D. Hruschka & Hadley, 2016; D. Hruschka et al., 2014).

As with weight-for-height, a universal reference for stunting may also seriously underestimate growth faltering in some populations by conflating variation due to deprivation (i.e. accrued HAZ) with variation in basal HAZ (Hruschka & Hadley, 2016).  Using Haiti and India as hypothetical examples, Figure 1 illustrates how a relatively deprived population with tall basal HAZ may seem just as healthy as a relatively privileged population with shorter basal HAZ (Figure 1).

Determining the relative contributions of these two components to HAZ has implications for research in a range of fields that rely on height-for-age as a measure of healthy growth.  This includes current global efforts to monitor undernutrition (Annan, 2018; Osgood-Zimmerman et al., 2018) and to track progress towards international development goals (De Onis et al., 2013; Stevens et al., 2012), as well as studies of the impact of culture (McDade et al., 2007), kinship and family dynamics (Gibson & Mace, 2005), ethnic disparities  (Lawson et al., 2014), anti-poverty and development programs (Behrman & Hoddinott, 2005) on well-being. For example, the use of HAZ scores to compare undernutrition across major world regions has spurred a substantial literature on an "Asian Enigma", whereby children in South Asian countries have unusually low HAZ scores despite the country's relatively high incomes (Deaton & Drèze, 2009; D. J. Hruschka & Hackman, n.d.; Jayachandran & Pande, 2017; Panagariya, 2013).

Due to the frequent use of HAZ as a proxy for stunting and undernutrition, the measure (HAZ) and concept (stunting due to undernutrition) are often used synonymously in these literatures(Frongillo et al., 2019). Thus, identifying how much each of the two components contribute to observed height-for-age in different populations should improve interpretation of HAZ as a measure of stunting and deprivation across a range of fields.

Here we describe an approach to partitioning population HAZ into these two components, using Demographic and Health Survey data on 1,522,252 children from 70 low- and middle-income countries (1990-2018). In this study, we focus on HAZ as opposed to other potential metrics due to is frequent use in global health monitoring(Leroy, Ruel, Habicht, & Frongillo, 2015).  First, we model children's HAZ as a function of a wide range of environmental variables that have been shown to influence child growth, including economic resources, maternal characteristics, disease burden, nutrition, healthcare access, and hygiene and sanitation at multiple levels (e.g., household, community, and country).  Comparing both linear and sigmoid models, we demonstrate that at extreme levels of deprivation, mean HAZ reaches a minimum (or basal) level below which further deprivation has little relationship with height.   We further show that this basal HAZ level differs substantially and reliably between countries, and these differences are uncorrelated with infant mortality.  These findings support an interpretation of basal HAZ as the component of variation in HAZ that is independent of environmental inputs and is unrelated to mortality.   By contrast, we show that the remaining component of HAZ (i.e. accrued HAZ) is strongly correlated with population estimates of infant mortality, indicating that accrued HAZ is the component of HAZ that reflects the standard definition of stunting as compromised height due to environmental insults.  Finally, we then examine how adjusting stunting estimates for such basal differences in HAZ can dramatically change the ranking of countries in terms of their vulnerability to growth faltering.

**METHODS**

<u>Sample:</u> Demographic and Health Surveys are nationally representative household surveys that collect information on a range of health and socioeconomic indicators. We used data from 225 surveys from 1990 to 2018 from 70 countries.  Early surveys only measured children of sampled women (rather than all children in the household).  For comparability across surveys, we limit all analyses to children of women selected for the survey.  Of 1,649,692 children ages 0-59 m who were eligible and present for height measurement, 0.16% were too sick for measurement, 1.7% were refusals, and 3.7% were missing height measurements for other reasons.  Of the 1,558,397 with height measurements, 2.3% had height-for-age z-scores that were > 6 SD or < -6 SD  (See SM Table S1 for survey-specific statistics, Figure S1 for survey-level distribution of proportion of extreme values).  The remaining 1,522,564 cases were used to plot mean height-for-age against household wealth per capita (Figure 2; 0-5 months n = 171,830, 6-11 months n = 169,977, 12-35 months n = 631,832, and 36-60 months n = 548,925).  Given the low

sensitivity of HAZ to environmental inputs among children ages 0-5 month, we then limit remaining analyses to children > 6 months (Wright, 2000).

DHS datasets frequently only collect certain variables (e.g. antenatal visits, maternal iron supplementation, child feeding) from most recently born children and/or children < 36 months. Thus, we focus our main analyses to most recently born children who are 12-35 months with available data on key covariates (220 surveys, N=514,675). However, we also assess how robust basal estimates are to: (1) analyzing the sample of older children 36-59 months (N=236,318), (2) analyzing the sample of younger children 6-11 months (N=163,428), and (3) analyzing a sub-sample of 12-23 month old children with data on recent intake of animal source proteins (N=244,566). To examine any gender differences in the effects of the explanatory variables on HAZ we run separate models for boys and girls.

Dependent variable: *Height-for-Age z-scores (HAZ).* We used the WHO SPSS anthro macros (http://www.who.int/childgrowth/software/en/) to estimate HAZ for all children based on height, age, sex, and whether the measurement was made standing or lying. We follow WHO guidelines to exclude children with implausible anthropometric values of +/- 6 SD (WHO, 2006; Zuguo & Grummer-Strawn, 2007).

Explanatory variables: The explanatory variables represent sources of influence on childhood growth, ranging from resource access and prenatal and postnatal care to hygiene, nutrition, and infectious disease exposure (Headey, Hoddinott, & Park, 2016). Additional information about the specific variables are available in the supplemental materials (Variable Description and Table S2 Variable Descriptives by Region).

*Child-level:* We include child age with a change in slope at 24 m (Leroy, Ruel, Habicht, & Frongillo, 2014; Leroy et al., 2015; Shrimpton et al., 2001), birth order dummy coded as first born, second born, and later born (Jayachandran & Pande, 2017), and dichotomous variables for > 3 antenatal visits, facility birth, > 7 vaccinations, mother's iron supplementation during pregnancy.  For a sub-sample of 12-23 month old children, we also include recent consumption of non-dairy animal source proteins and dairy-based animal source proteins (Baten & Blum, 2012; Grasgruber, Cacek, Kalina, & Sebera, 2014).

*Mother-level:* We include a linear and quadratic term for mother's age (centered at 30 y), parity top-coded at 12 children, a dichotomous variable for literacy, and dummy coded education (no education, primary, secondary, and post-secondary).

*Household-level:* We include urban residence, household open defecation, and log-transformed absolute wealth estimates (AWE) based on assets, housing construction and service access. The latter measure facilitates comparisons of household wealth both within a country across different survey years, as well as across survey populations, in absolute units— 2011-constant international dollars with purchasing power parity (D. Hruschka, Gerkey, & Hadley, 2015).

*Cluster-level.* Cluster-level variables capture ecological factors beyond the households at the level of primary sampling units.  These units are neighborhoods or clusters of households selected for 2[nd] level sampling in demographic and health surveys, with clusters usually, but not always, representing roughly 20 households (mean = 21.6, SD = 16.8, range = 1 to 844). We include cluster level proportions of open defecation with a spline below 0.30  (Headey et al., 2016).   To assess infectious disease exposure, we also include the proportion of households in both a cluster and a country's first-level administrative district with a child who experienced diarrhea in the last week.

*Study Year.* We include a variable indicating year since 1990, our earliest set of surveys, to capture any potential increases in HAZ over time that are not captured by our explanatory variables.

<u>Validation Measures:</u> To validate our estimates of the two components of HAZ (basal and accrued HAZ), we assess the predicted associations between these estimates and country/year-level estimates of all cause infant mortality.

*Infant Mortality.* Estimates of infant mortality rates for the survey year were taken from the World Bank Indicators website (https://data.worldbank.org/indicator/SP.DYN.IMRT.IN).  These estimates were developed by the UN Inter-agency Group for Child Mortality Estimation at childmortality.org.

**ANALYSIS PLAN**

<u>*Modeling HAZ as a function of environmental inputs.*</u>  Past research indicates that a sigmoid model fits the relationship between human height and environmental inputs better than a linear model (D. Hruschka, Hackman, & Stulp, 2019) .  As opposed to a linear model, a sigmoid model exhibits: (1) a nadir below which there is no longer any reduction in height with declining resources and (2) declining marginal returns to height with increasing resources.

To visually assess bottoming out of the relationship between resources and HAZ across the full sample, we first plot the mean HAZ among all children in four age categories - 0-5 months, 6-11 months, 12-35 months and 36-60 months - across 16 categories of household wealth per capita (Figure 2).

We then formally assess the fit of both linear (equation 2) and sigmoid (equation 3) models predicting HAZ as a function of environmental variables ($\sum \beta_k X_k$) as well as a population-specific intercept that can vary between populations ( $d_{pop}$).

$$HAZ = \sum \beta_k X_k + d_{pop} + \varepsilon_{pj} \qquad (2)$$

$$HAZ = \frac{a}{1+e^{(c-\sum \beta_k X_k)}} + d_{pop} + \varepsilon_{pj} \qquad (3)$$

In both cases, $\sum \beta_k X_k$ is a linear combination of the individual, household, and cluster level environmental inputs representing increasing resources. We refer to this as a *resource score* for a child. The sigmoid curve relating the resource score and HAZ involves two additional parameters. Parameter *a* captures the height difference between the nadir and the plateau of the sigmoid curve and reflects the maximum gains in height that a population can achieve through improving environmental conditions. Parameter *c* captures the inflection point of the sigmoid curve.

In the sigmoid model, $d_{pop}$ is the expected HAZ value for children in population *pop* at the nadir of the sigmoid curve (e.g., the most extreme level of deprivation). In the sigmoid model, this has a natural interpretation as basal HAZ. In other words, $d_{pop}$ is a population-specific starting point from which a population can increase as it enjoys better nutrition, lower disease burden and other improved environmental inputs.

To assess the possibility that parameter *a* in the sigmoid model varies between region, we permit parameter *a* to differ for sub-Saharan Africa and for South and Southeast Asia by including dummy fixed effects (using children from other regions as the reference category).

We used linear and non-linear mixed effects models in R (lme and nlme) to fit the linear and sigmoid models respectively (Bates et al., 2017). These included nested random effects of survey within country. For country-specific basal HAZ values (i.e., the $d_{pop}$ parameter), we use the conditional modes of the random effects for each country estimated using REML (Bates, 2010, Bates et al., 2017; Faraway, 2016; Zuur, Ieno, Walker, Saveliev, & Smith, 2009). To compare fits between linear and sigmoid models we use Akaike Information Criteria estimated with maximum likelihood.

In case wealth may have smaller effects among rural subsistence farmers than urban residents, we tested an interaction between wealth and urban residence. However, these interactions were not significant so the interactions were not retained in the model.

_Estimating accrued HAZ._ We use the basal HAZ values to estimate mean accrued HAZ for each survey (accrued HAZ = survey mean HAZ – basal HAZ).

_Assessing robustness of basal HAZ to different samples and model specifications._ If basal HAZ represents relatively stable country-level differences in HAZ that are independent of environmental inputs, then these estimates should not vary substantially between surveys from the same country or between sexes or age groups from the same survey.  We assess the reliability of survey-year estimates of basal HAZ as indicators of country-level basal HAZ by estimating the proportion of between-survey variation in basal HAZ estimates that is due to between-country variation in basal HAZ estimates.  We assess the consistency of sex-specific estimates of basal HAZ by assessing the correlation between country-level basal HAZ estimated separately for girls and for boys. We also assess how robust these estimates are when based on: (1) children of different ages (36-59 m and 6-11 m vs. 12-35 m), (2) on models including a more restricted sample (12-23 m) with data on recent intake of animal source proteins, and (3) on models permitting the effect of birth order to be greater among Hindu families(Jayachandran & Pande, 2017).  Based on the high consistency of country-specific basal HAZ estimates across surveys, between sexes, between age groups, and with alternative model specifications, we use country-specific estimates for 12-35 m old children for further analyses.

_Assessing validity of accrued and basal HAZ._ While we treat basal HAZ as a country-level concept in this paper, accrued HAZ and observed HAZ can change within a country over time as environmental inputs change.  We assess the validity of the decomposition of HAZ into country-level basal HAZ and survey-level mean accrued HAZ estimates as follows.  First, we compare these estimates with contemporary all-cause infant mortality. If basal HAZ is independent of health outcomes, then we expect little correlation between basal HAZ and infant mortality. By contrast, the mean accrued HAZ for a survey is expected to capture the portion of HAZ that is sensitive to resource inputs and also relevant to healthy development. Thus, accrued HAZ should show stronger negative associations with infant mortality than observed HAZ, since we are partialling out the variance in HAZ due to basal differences.

*How do estimates of stunting change when using accrued HAZ versus observed HAZ.* After establishing that basal HAZ is stable and independent of environmental inputs and infant mortality and that accrued HAZ captures the component of HAZ that is associated with environmental inputs and infant mortality, we explore how estimates of stunting change when using accrued versus observed HAZ. First, we examine how rankings of populations change when comparing accrued HAZ versus observed HAZ.

We conduct a second exercise based on stunting prevalence because researchers commonly use WHO cutoffs to estimate and compare stunting prevalence across populations. That said, there are important caveats when using and interpreting stunting prevalences based on WHO cutoffs. First, the 3 SD cutoff is biologically arbitrary (Leroy & Frongillo, 2019; Perumal et al., 2018). Second, stunting prevalence is more appropriately interpreted as an indicator of deprivation for the entire population (not only among those classified as stunted)(Roth et al., 2017).

Given these and related caveats, we conduct the second exercise to illustrate further issues with using a universal standard. First, we need to identify a threshold for accrued HAZ that is equivalent to the WHO threshold for severe stunting. For example, an Indian child with an observed HAZ below -3 SD would be counted as severely stunted. Returning to equation 1 that relates accrued, basal and observed HAZ, we can derive a threshold of accrued HAZ which would be equivalent to the WHO cutoff for severe stunting. As illustrated in equation 4, this would be $-3 - bHAZ_{India}$

$$bHAZ_{India} + aHAZ < -3 \qquad (3)$$

$$aHAZ < -3 - bHAZ_{India} \qquad (4)$$

We refer to this quantity as the aHAZ threshold for severe stunting. Because the aHAZ threshold depends on basal HAZ, it will vary by the reference population we use to calculate it. In this paper, we use India as a reference population for several reasons. First, India has high rates of childhood stunting (Martorell & Young, 2012). Second, there is a long history of assessing malnutrition using childhood anthropometrics among Indian children (Nandy, Irving, Gordon, Subramanian, & Smith, 2005; Radhakrishna & Ravi, 2004). Third, India's capital was one of the sites used in the creation of the WHO standards (De Onis et al., 2004). Finally, India is the most populous of countries in the dataset and constitutes a large part of the total sample (N=328,719; 22% of total sample).

By choosing India as the reference population, we are assuming that the WHO cutoffs are appropriate for assessing stunting based on HAZ scores in India. Changing the reference population will uniformly up- or down-shift stunting thresholds for all countries depending on the reference population's basal HAZ level. However, it will not change the relative ranking of those cutoffs between countries. Using the accrued HAZ threshold derived from India, we use design-weighted sample proportions to estimate the prevalence of severe stunting across the full-range of countries in our analyses. We then compare these to stunting prevalence estimates based on the original WHO cutoffs.

**RESULTS**

*Modeling HAZ as a function of environmental inputs.* Plotting the mean HAZ by wealth illustrates the bottoming out of the relationship between material wealth and linear growth (Figure 2). These preliminary results show that HAZ scores are somewhat sensitive to household wealth for children 6-11 m, but most sensitive to household wealth for children 12-59 m and that the effect is similar for children 12-35 m and 36-59 m. Notably, HAZ scores are much less sensitive to increasing wealth among children 0-5 m, which is consistent with prior research on age-specific sensitivity of cross-sectional HAZ to environmental inputs (Wright, 2000).

Consistent with the well-known and substantial effect of resources on HAZ, both sigmoid (Table 1, Figure 3) and linear (SM Table S3) models showed significant and substantial effects of a range of environmental inputs on child HAZ. For both sexes, there were negative effects of neighborhood-level diarrhea and open defecation, confirming a flattening out of the effect of open defecation above 30% of households in the neighborhood (Headey et al., 2016). There were strong positive associations with household wealth, measures of health care access, and maternal education. Supplemental analyses also showed strong positive associations with recent intake of animal source proteins (SM Table S4). Maternal age showed a curvilinear effect on HAZ, with lower HAZ among children born to younger and older mothers. Consistent with prior research, birth-order showed a negative association with HAZ (Jayachandran & Pande, 2013), and child age showed a strong negative effect between 12 and 23 months, and a relatively flat effect after 23 months (Leroy et al., 2014). Finally, there is a slight, but statistically significant effect, of survey year.

Importantly, the sigmoid model relating resources and HAZ (which permits bottoming out of the effect of environmental inputs on HAZ) provides a much better fit to the data than linear models that do not permit such bottoming out ($\Delta$AIC = -104.0 for boys and -144.1 for girls). Moreover, the country-specific estimates of the basal level where HAZ bottoms out (i.e., basal HAZ) also show substantial variation (Figure 3). Specifically, the country-level basal HAZ estimates showed a full range of 1.3 SD across all countries for boys and 1.5 SD for girls (SM Table S5).

*Assessing robustness of basal HAZ estimates.* Country-level basal HAZ estimates were robust across a range of specifications, including survey-versus country-level estimates, boys versus girls, older versus younger children, as well as estimates derived from models including additional covariates and potential interactions. First, country-level bHAZ estimates are strongly correlated with study-level bHAZ estimates for both boys (r = 0.95) and girls (0.96) (Figure S2). These results indicate that country-level estimates are reliable and justify our focus on country-level estimates of basal HAZ rather than estimates for a specific survey year. Additionally, country-level estimates of basal HAZ were highly correlated across genders (boys vs. girls r=0.93) and across age groups (12-35 m boys vs. 36-59 m versus 12-35 m boys, r=0.81; 36-59 m versus 12-35 m girls r=0.81). Figure 4 shows correlation of estimates for boys and girls. Finally, estimates based on alternative model specifications also were strongly correlated with estimates from the main model (dietary variables for 12-23 boys r= 0.92 and girls r=0.96; Hindu-specific birth order (boys r = 0.99, girls r = 0.98).

*Assessing validity of accrued and basal HAZ.* As expected, mean observed HAZ has a strong negative correlation with measures of infant mortality (Girls r=-0.60, Boys r=-0.56). By contrast, basal HAZ estimates show little to no association with infant mortality (Girls r=-0.07, Boys r=-0.20), supporting the interpretation of basal HAZ as a measure largely unrelated to key environmental inputs or mortality risks (Figure 5). While basal HAZ shows low to non-existent associations with population mortality risk, accrued HAZ shows strong negative associations with infant mortality (Girls r=-0.69, Boys r=-0.64). In all cases, accrued HAZ showed stronger association with infant mortality than the standard observed HAZ measures. This suggests that accrued HAZ captures the component of observed HAZ that reflects standard definitions of stunting as compromised growth with negative health consequences.

_How do estimates of stunting change when using accrued HAZ versus observed HAZ._  Ranking population-level stunting by accrued HAZ instead of observed HAZ substantially changed the relative ranking for numerous surveys.  Table 2 shows the average change in rankings for 16 countries that climbed highest in the rankings.  These include 11 countries from West Africa, 3 from Latin America and the Caribbean (Haiti, Dominican Republic, and Nicaragua), as well as Morocco and Turkey.

[Table 2]

We arrive at similar results when re-estimating prevalence of severe stunting for each of these surveys based on accrued HAZ.  Using basal HAZ for our selected reference population (i.e., India), we estimate the accrued HAZ threshold as 1.26 SD for boys and 3.73 SD for girls. Any child with accrued HAZ less than these values would be classified as severely stunted.  Figure 6 compares the prevalence of severe stunting estimated with observed HAZ (using standard WHO cutoffs) and with accrued HAZ (using the accrued HAZ threshold).  Nearly all countries have a higher estimated prevalence of severe stunting using the accrued HAZ threshold derived from India.  The few exceptions are Pakistan, Guatemala, and East Timor which have even lower basal HAZ estimates than India.

Consistent with the findings using observed and accrued HAZ, populations from West Africa saw large upward shifts in prevalence estimates when using the accrued HAZ threshold.  These ranged from an average upward shift of 0.14 to 0.21 (Table 2).

**DISCUSSION**

Consistent with well-established research on child growth, we identify substantial associations between a range of environmental inputs—nutritional, disease burden, socioeconomic resources—and children's HAZ scores.  However, even after accounting for these diverse factors, there remain substantial between-country differences (a range of 1.5 standard deviations) in children's height.

Partitioning this variation in observed HAZ into one component that is sensitive to environmental inputs—accrued HAZ—and another that is unrelated to environmental inputs—basal HAZ—leads to several observations.  First, between-country differences in basal HAZ are robust across samples based on sex and age. Consistent with a two-component model, these differences in basal HAZ are largely unrelated to population estimates of infant mortality. Importantly, our analyses include variables, such as birth order, sanitation and recent consumption of animal source proteins, which have recently been proposed to account for puzzling regional differences between sub-Saharan Africa and South Asia (Baten & Blum, 2012; Grasgruber et al., 2014; Jayachandran & Pande, 2017; Spears, 2018)..  While these variables are important predictors of HAZ, they do not substantially attenuate the estimated between-country differences in basal HAZ.

In contrast to basal HAZ, population estimates of accrued HAZ show strong associations with infant mortality, and these associations are even stronger than those between observed HAZ and infant mortality. Taken together, these findings suggest that estimates of accrued HAZ are the quantities of interest when comparing populations in terms of environmental deprivation and health outcomes.

Using accrued HAZ rather than observed HAZ to prioritize survey populations by stunting risk substantially revises rankings, with West African countries witnessing the most substantial upward adjustments in relative stunting (Table 2).  When population differences are framed in terms of proportion of children suffering from severe stunting, estimates based on accrued HAZ threshold are also substantially larger in these surveys compared to estimates based on WHO standard cutoffs (0.14 to 0.21 greater prevalence). In many cases, this amounts to a doubling or tripling of prevalence estimates for severe stunting.

The WHO growth standards reports hold the position that since ~90-95% of the variance in HAZ falls within populations, any between-group differences can be effectively ignored (De Onis, 2006; Habicht, Yarbrough, Martorell, & Malina, 1974; WHO Multicentre growth reference study group, 2006). The between- and within-population variance reported here falls within the ranges reported by the MGRS (Buuren & Wouwe, 2008). However, we also demonstrate that even a small amount of between-population variation in basal HAZ can lead to substantial underreporting of stunting in specific world regions depending on the reference population one uses.

The WHO standard creates a powerful policy message that when needs are met, children grow very similarly regardless of where they live or their ethnic background. It is true that increasing economic resources and nutrition is associated with increasing linear growth across these samples. However, populations also appear to differ (often substantially) in their basal levels. By assuming that all children have the same starting point at extremes of deprivation, universal cutoffs may neglect children in countries and regions where healthy HAZ tends to be higher.

Consider a comparison of Haiti with either Guatemala or India. Between 2015 and 2017, Haiti's mean observed HAZ was substantially higher than either of the other country's observed HAZ (-0.94 vs. -1.89 for Guatemala and -1.42 for India). The estimated prevalence of severe stunting was also much lower in Haiti (8% vs. 18% for Guatemala and India). By universal standards applied to observed HAZ, Haitian children would seem to be enjoying far better conditions for growth than their counterparts in the other two countries. However, Haiti had a much lower GDP per capita (1653 constant 2011 international dollars vs. 7293 and 5743) and much higher infant mortality rates (53.9 per 1000 live births vs. 24.7 and 35.3). Indeed, there are few indicators by which one would rank Haiti better than Guatemala or India for child growth during this time period. Making the comparison by accrued HAZ solves this conundrum, reversing

Haiti's puzzling ranking in terms of stunting prevalence (24% for Haiti vs. 16% Guatemala and 18% for India).

Future Directions

The current analyses raise a number of questions for future study. First, we chose country of residence as the basis for grouping individuals into populations because of the availability of country-level indicators and the use of countries as a common unit for global health monitoring. However, finer-grained variation in basal HAZ may be achieved by examining subgroupings within countries based on subdistrict of residence and ethnolinguistic affiliation (D. Hruschka, Hadley, et al., 2015).

Second, to further calibrate accrued HAZ as an indicator of undernutrition and to assess its improvement over observed HAZ for tracking individual and population well-being, we need more direct markers of economic, nutritional, and health care resources. A number of explanatory variables used in our model are not measured at the level of the household, but at the cluster or subdistrict level. Finer-grained measurements of infectious disease exposure, dietary quality and diversity, and health care access at the household level would help refine these estimates of basal HAZ across populations.

The models estimate similar associations between environmental resources and HAZ for boys and girls. However, the models do differ substantially in the estimated height of the sigmoid curve (4.05 SD for boys vs. 6.76 SD for girls, Table 1). Future work should identify if this is due to sex differences in the sensitivity to environmental inputs or other potential reasons. Moreover, basal HAZ estimates from 12-35 m children are highly consistent with those from 6-11 m and 36-59 m children. That said, we might expect population specific basal growth trajectories that deserve further exploration.

In this study, we followed WHO guidelines in excluding the cases with HAZ magnitudes greater than 6 SD.  Future work should examine the sensitivity of these findings to changing these cutoffs (Roth et al., 2017).  We also focused on height-for-age z-scores because they are the most commonly used metric for assessing stunting and growth faltering internationally. However, following the same approach with metrics in absolute units (e.g. centimeters above a reference median) should shed further light on partitioning height variation into population-specific starting points and resource-driven increases (Leroy et al., 2015).

Finally, comparing HAZ with other resource-sensitive development indicators across populations may also improve our understanding of the meaning of HAZ as a measure of deprivation and vulnerability. Measures of cognitive, motor, and brain development (Kar, Rao, & Chandramouli, 2008; Tarleton et al., 2006) could serve as alternative indicators of adequate development. Like height, cognitive development is sensitive to resource inputs, indeed showing strong correlations with HAZ  (Spears, 2012). Such additional measures would help determine when variation in height reflects deprivation and vulnerability and when it does not.

**CONCLUSION**

This work adds to a growing body of literature showing the importance of incorporating population variation in body size when using anthropometrics to assess health globally (D. Hruschka & Hadley, 2016).  The approach taken here opens up the possibility of assessing population differences in growth without the restricted sampling of only those children raised in environments deemed ideal (Karra et al., 2017). Such an approach would add to our understanding of the full range of human childhood growth (Christesen et al., 2016; Natale & Rajagopalan, 2014).  It also has the potential to identify variation in the meaning of HAZ as a measure of undernutrition in different worldwide populations, and to identify those populations that might be missed by universal standards for normal growth.

**Acknowledgements:** The authors would like to thank Craig Hadley, Peter Rohloff, Ben Trumble, Gert Stulp, Rebecca Sear, and Alexandria Drake for constructive comments on the manuscript.  DJH acknowledges support from the National Science Foundation grant BCS-1150813, jointly funded by Programs in Cultural Anthropology, Social Psychology Program and Decision, Risk, and Management Sciences, and BCS-1658766, jointly funded by Programs in Cultural Anthropology and Methodology, Measurement and Statistics, and support from the Virginia G Piper Charitable Trust through an award to Mayo Clinic/ASU Obesity Solutions. The funder had no role in study design, data collection and analysis, decision to publish, or preparation of the manuscript.

**Data Availability Statement:**  The data used for this study are publicly available from The DHS Program (https://www.dhsprogram.com/) and The World Bank (https://data.worldbank.org/indicator).

**Figure Legends**

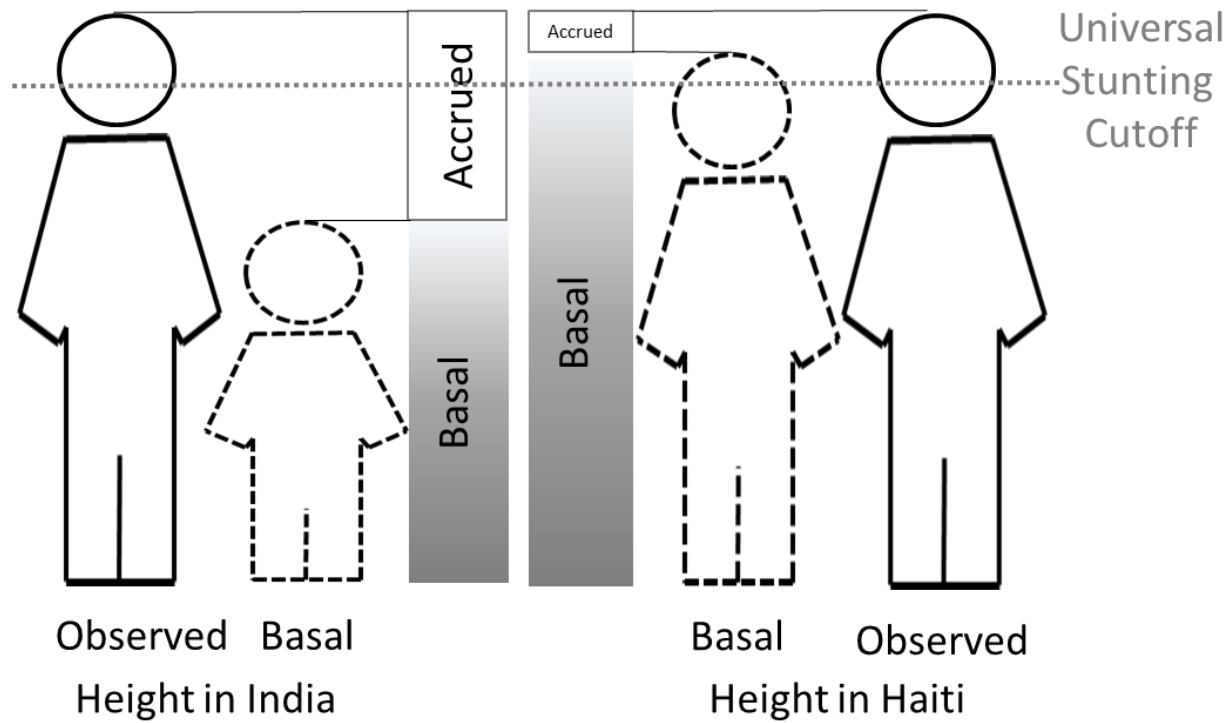

**Figure 1. Hypothetical relationship between basal, accrued, and observed height in two populations**. In this situation, Indian and Haitian children have similar observed heights, and both are above the universal stunting cutoff. However, because they had different basal starting points, the Haitian children have experienced much smaller gains from environmental inputs (accrued height) than Indian children.

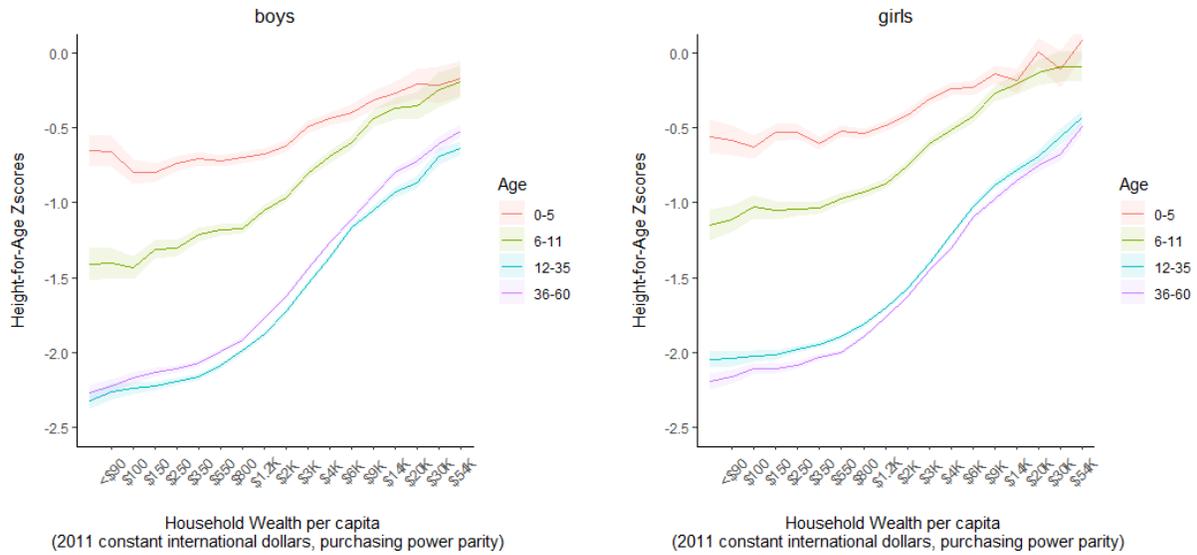

**Figure 2. Mean height-for-age z-scores by estimated household wealth per capita for the full sample (by age category).** Shaded regions represent 95% confidence intervals around the mean HAZ for a given wealth category. The x-axis reflects the mean wealth of the binned wealth category.

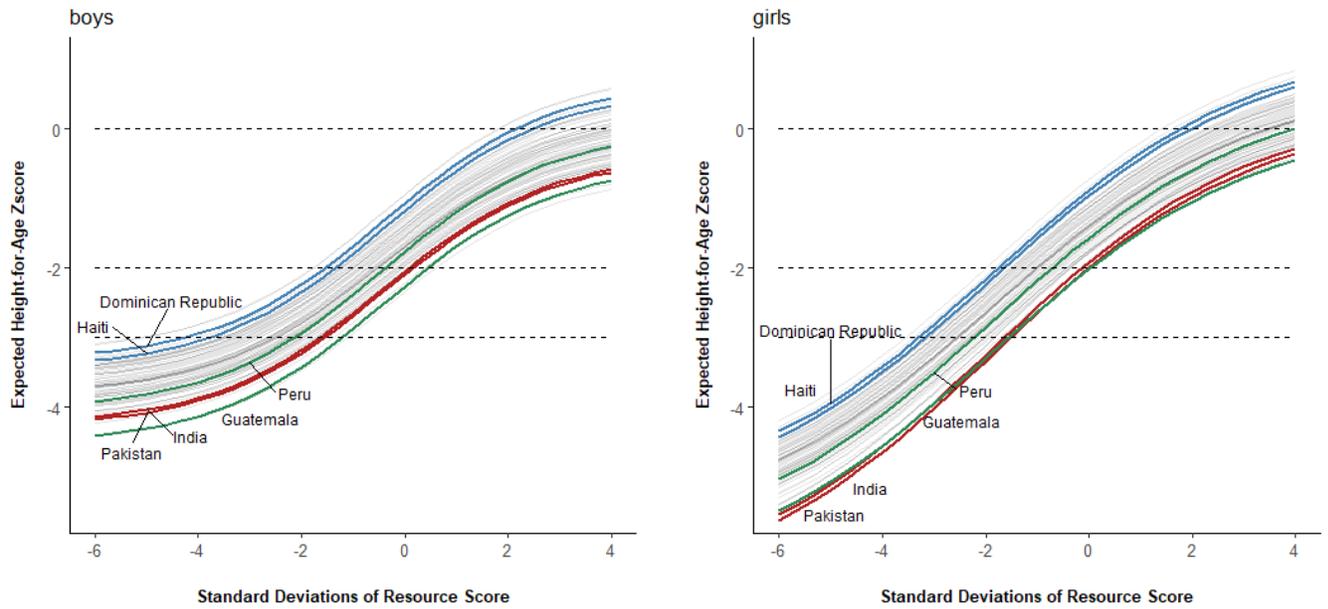

**Figure 3. Expected HAZ as a function of resource-driven increases and country-specific basal differences.** Country-specific estimates for HAZ plotted over increase resource scores. The dashed line represents the WHO cutoffs for moderate and severe stunting. Each country is represented by one curve, with highlighted lines for Haiti & Dominican Republic (blue), Pakistan & India (red), Guatemala & Peru (green).

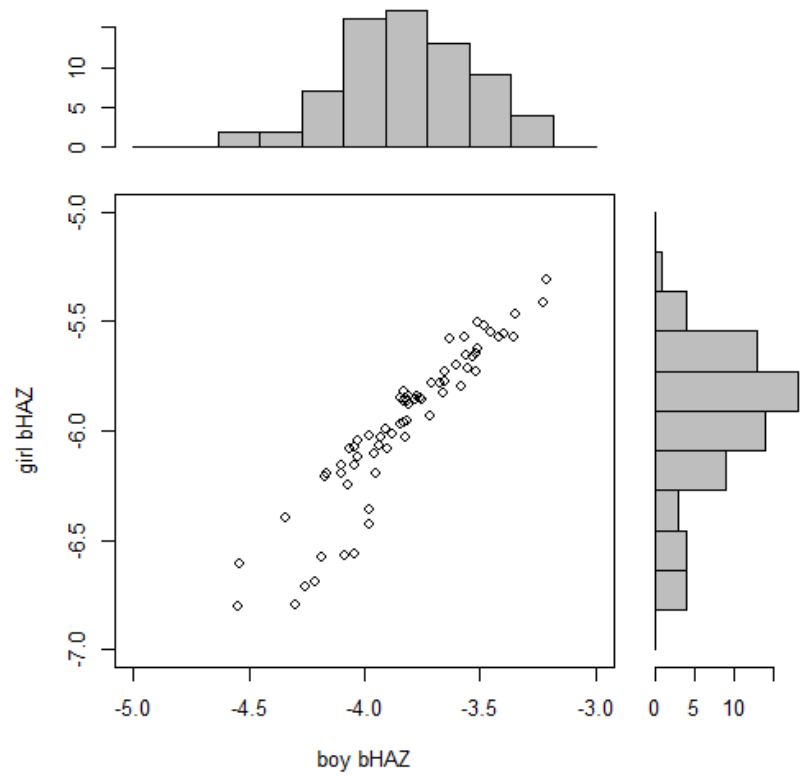

**Figure 4. Country-specific basal HAZ estimates for boys and girls.**

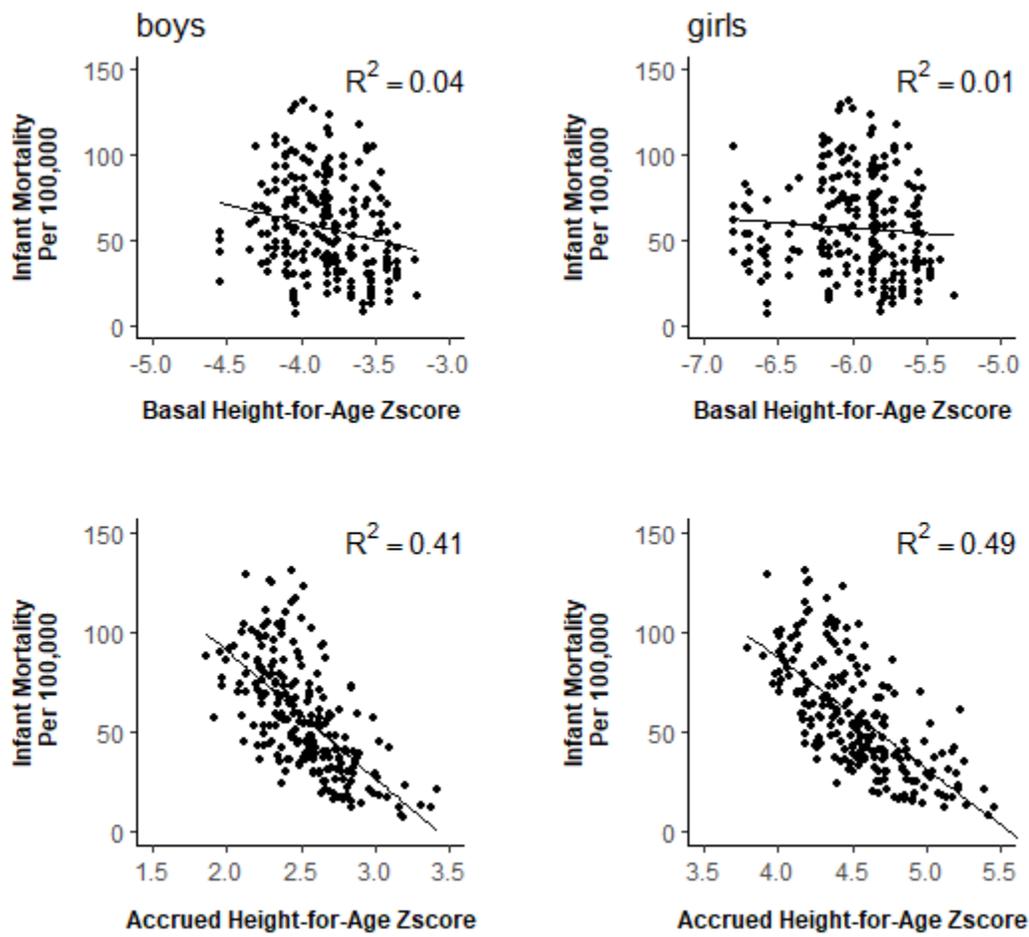

**Figure 5. Associations between the two components of height-for-age (country-level basal HAZ and survey-level accrued HAZ) and survey-year estimates of Infant Mortality.**

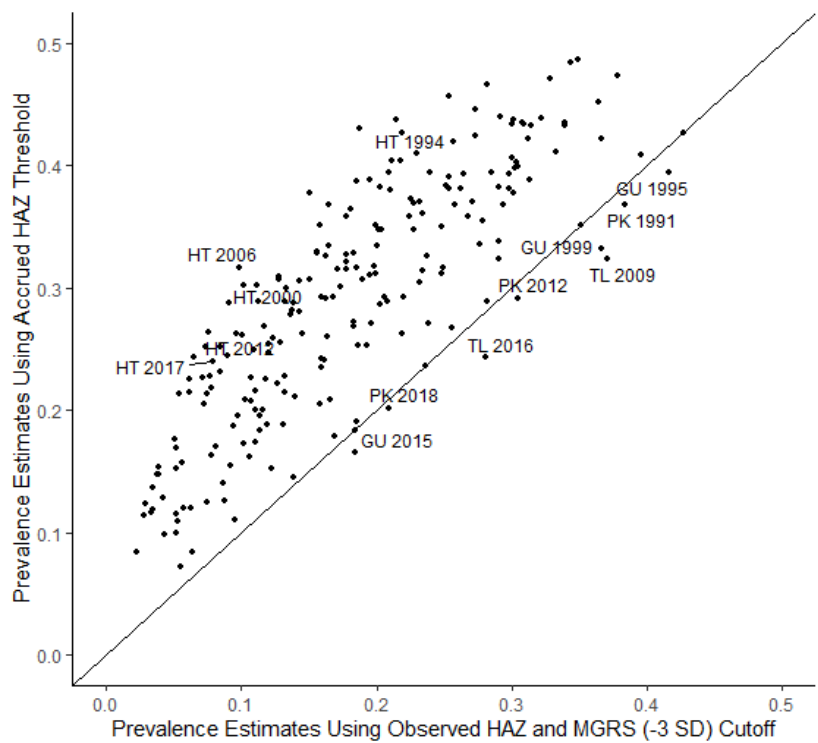

**Figure 6. Comparing severe stunting prevalence estimates based on universal MGRS cutoffs and accrued HAZ (with India as reference population).** Haiti surveys and all data points below the line of equality are labeled. GU – Guatemala; TL – East Timor; PK – Pakistan.

Table 1.  Non-linear mixed effects models for boys and girls (12-35 m)

| Predictors | Boys 12-35 | | Girls 12-35 | |
|---|---|---|---|---|
| | Estimates | CI | Estimates | CI |
| *Sigmoid Curve Parameters* | | | | |
| Increase from Lower Asymptote | 4.05 * | 3.43 – 4.68 | 6.76 * | 4.10 – 9.42 |
| Inflection point (c) | 1.13 * | 0.82 – 1.45 | 0.06 | -0.48 – 0.60 |
| Lower Asymptote (d) | -3.68 * | -4.12 – -3.24 | -5.85 * | -8.09 – -3.61 |
| *Child Variables* | | | | |
| Child Age | 0.00 | -0.00 – 0.01 | -0.01 * | -0.01 – -0.00 |
| Child Age > 24 | -0.05 * | -0.06 – -0.04 | -0.03 * | -0.03 – -0.02 |
| Facility Birth | 0.13 * | 0.10 – 0.16 | 0.08 * | 0.06 – 0.11 |
| Facility Birth-Missing | 0.05 | -0.04 – 0.14 | 0.04 | -0.01 – 0.10 |
| > 3 Antenatal Visits | 0.13 * | 0.11 – 0.16 | 0.08 * | 0.05 – 0.11 |
| Antenatal Visits-Missing | 0.10 * | 0.04 – 0.16 | 0.08 * | 0.04 – 0.13 |
| > 7 Vaccinations | 0.03 * | 0.02 – 0.05 | 0.03 * | 0.02 – 0.05 |
| Birth Order = 2 | -0.06 * | -0.09 – -0.04 | -0.05 * | -0.07 – -0.03 |
| Birth Order > 2 | -0.12 * | -0.15 – -0.08 | -0.10 * | -0.13 – -0.06 |
| *Mother Variables* | | | | |
| Number of Children | -0.05 * | -0.06 – -0.04 | -0.03 * | -0.04 – -0.02 |
| Literacy | 0.08 * | 0.06 – 0.11 | 0.04 * | 0.02 – 0.06 |
| Primary Education | 0.05 * | 0.02 – 0.08 | 0.04 * | 0.02 – 0.06 |
| Secondary Education | 0.23 * | 0.18 – 0.28 | 0.15 * | 0.10 – 0.21 |
| Higher Education | 0.51 * | 0.40 – 0.61 | 0.35 * | 0.23 – 0.47 |
| Mother's Age | 0.26 * | 0.22 – 0.31 | 0.17 * | 0.12 – 0.23 |
| Mother's Age Squared | -0.06 * | -0.08 – -0.04 | -0.03 * | -0.04 – -0.02 |
| *Household Variables* | | | | |
| Absolute Household Wealth | 0.17 * | 0.14 – 0.19 | 0.10 * | 0.06 – 0.13 |
| Open Defecation-Household | -0.04 * | -0.06 – -0.02 | -0.01 | -0.03 – 0.00 |
| Urban | 0.04 * | 0.02 – 0.06 | 0.03 * | 0.01 – 0.04 |
| *Cluster, Subdistrict, & Survey* | | | | |
| Open Defecation-Cluster | -0.21 * | -0.30 – -0.12 | -0.18 * | -0.25 – -0.10 |
| Open Defecation > 0.30 | 0.24 * | 0.12 – 0.36 | 0.19 * | 0.10 – 0.28 |
| Diarrhea-Cluster Level | -0.26 * | -0.34 – -0.18 | -0.13 * | -0.19 – -0.08 |
| Diarrhea-Subdistrict Level | -0.57 * | -0.76 – -0.38 | -0.30 * | -0.45 – -0.16 |
| Survey Year | 0.02 * | 0.01 – 0.02 | 0.01 * | 0.01 – 0.01 |
| *Region Deviations* | | | | |
| sub-Saharan Africa Δd | -0.14 | -0.31 – 0.02 | -0.06 | -0.29 – 0.17 |
| South & Southeast Asia Δd | -0.52 * | -0.76 – -0.28 | -0.80 * | -1.11 – -0.50 |
| sub-Saharan Africa Δa | -0.25 * | -0.41 – -0.10 | -0.20 | -0.46 – 0.05 |

| South & Southeast Asia Δa | -0.13 | -0.28 − 0.03 | 0.31 | 0.02 − 0.59 |
|---|---|---|---|---|
| Observations | | 263650 | | 251025 |
| N_countries | | 70 | | 70 |
| N_surveys | | 220 | | 220 |

**Table 2. Change in survey rankings when using accrued HAZ versus observed HAZ** (16 countries with the largest increases). LAC = Latin America and Caribbean

| | region | # surveys | Mean increase in rank based on HAZ | | Ave. increase in prop. stunting all sexes |
|---|---|---|---|---|---|
| | | | boys | girls | |
| **Burkina Faso (BF)** | W. Africa | 4 | 30.5 | 39.3 | 0.17 |
| **Cote D'Ivoire (CI)** | W. Africa | 3 | 70.3 | 58.3 | 0.16 |
| **Dom. Republic (DR)** | LAC | 5 | 57.3 | 64.0 | 0.14 |
| **Ghana (GH)** | W. Africa | 5 | 35.2 | 41.8 | 0.16 |
| **Gambia (GM)** | W. Africa | 1 | 44 | 57 | 0.17 |
| **Guinea (GN)** | W. Africa | 3 | 77.7 | 69.0 | 0.18 |
| **Haiti (HT)** | LAC | 5 | 76 | 81.2 | 0.18 |
| **Liberia (LB)** | W. Africa | 2 | 46.5 | 80.5 | 0.17 |
| **Morocco (MA)** | N. Africa | 2 | 84 | 71.5 | 0.18 |
| **Mauritania (MR)** | W. Africa | 1 | 64 | 87 | 0.20 |
| **Nicaragua (NC)** | LAC | 2 | 66 | 48.5 | 0.16 |
| **Sierra Leone (SL)** | W. Africa | 2 | 54.5 | 59.5 | 0.16 |
| **Senegal (SN)** | W. Africa | 8 | 87.6 | 78.1 | 0.18 |
| **Sao Tome (ST)** | W. Africa | 1 | 73 | 101 | 0.18 |
| **Togo (TG)** | W. Africa | 2 | 86.0 | 101.0 | 0.21 |
| **Turkey (TR)** | W. Asia | 2 | 99.5 | 63.5 | 0.14 |

**For Online Publication**

**Appendix: Supplemental materials**

Tables S1. Survey-specific descriptives

Variable descriptions

Table S2. Variable descriptives by region (main model)

Table S3 linear models

Table S4 models with dietary

Table S5 country-level basal estimates

**Table S1**. Survey-specific descriptives

| study | eligible /present height | missing height | extreme HAZ | HAZ | HAZ (sd) | HAZ (boys) | HAZ (girls) | infant mortality | N-main model boys | girls | Accrued HAZ boys | girls | Severe Stunting observed HAZ | accrued HAZ |
|---|---|---|---|---|---|---|---|---|---|---|---|---|---|---|
| AL 2008 | 1574 | 50 | 72 | -0.71 | 2.31 | -0.75 | -0.67 | 12.3 | 253 | 227 | 2.84 | 5.13 | 0.20 | 0.27 |
| AL 2017 | 2755 | 244 | 19 | -0.39 | 1.55 | -0.40 | -0.38 | 7.8 | 510 | 424 | 3.18 | 5.42 | 0.05 | 0.12 |
| AM 2000 | 1595 | 56 | 6 | -0.82 | 1.39 | -0.78 | -0.85 | 26.5 | 274 | 200 | 2.73 | 4.87 | 0.04 | 0.15 |
| AM 2005 | 1351 | 51 | 27 | -0.53 | 1.65 | -0.55 | -0.51 | 20.7 | 245 | 194 | 2.96 | 5.21 | 0.05 | 0.15 |
| AM 2010 | 1432 | 26 | 27 | -0.86 | 1.79 | -0.80 | -0.92 | 16.1 | 271 | 225 | 2.72 | 4.81 | 0.11 | 0.20 |
| AM 2016 | 1703 | 100 | 22 | -0.22 | 1.68 | -0.15 | -0.28 | 11.9 | 289 | 263 | 3.37 | 5.45 | 0.04 | 0.10 |
| AO 2015 | 7065 | 502 | 145 | -1.54 | 1.56 | -1.50 | -1.58 | 58.2 | 996 | 966 | 2.58 | 4.66 | 0.19 | 0.25 |
| AZ 2006 | 2174 | 85 | 75 | -1.16 | 1.70 | -1.10 | -1.21 | 41.6 | 359 | 280 | 2.80 | 4.87 | 0.13 | 0.19 |
| BD 1996 | 5383 | 283 | 204 | -2.29 | 1.61 | -2.25 | -2.32 | 77.3 | 887 | 856 | 1.97 | 4.37 | 0.40 | 0.41 |
| BD 1999 | 6064 | 494 | 141 | -1.99 | 1.44 | -1.98 | -2.00 | 67.1 | 995 | 941 | 2.23 | 4.69 | 0.26 | 0.27 |
| BD 2004 | 6289 | 236 | 50 | -1.95 | 1.39 | -1.92 | -1.97 | 53 | 756 | 754 | 2.29 | 4.72 | 0.28 | 0.29 |
| BD 2007 | 5617 | 218 | 50 | -1.71 | 1.38 | -1.72 | -1.71 | 45.6 | 1015 | 999 | 2.50 | 4.98 | 0.17 | 0.18 |
| BD 2011 | 8138 | 269 | 129 | -1.65 | 1.46 | -1.67 | -1.63 | 36.9 | 1347 | 1368 | 2.55 | 5.06 | 0.18 | 0.19 |
| BD 2014 | 7455 | 321 | 102 | -1.53 | 1.37 | -1.52 | -1.54 | 31.5 | 1387 | 1312 | 2.69 | 5.15 | 0.14 | 0.15 |
| BF 1993 | 4858 | 264 | 97 | -1.40 | 1.69 | -1.34 | -1.45 | 98.8 | 810 | 777 | 2.37 | 4.33 | 0.20 | 0.38 |
| BF 1999 | 4852 | 78 | 132 | -1.70 | 1.81 | -1.66 | -1.74 | 92.9 | 789 | 829 | 2.05 | 4.04 | 0.28 | 0.47 |
| BF 2003 | 9252 | 454 | 277 | -1.59 | 1.94 | -1.49 | -1.68 | 86.3 | 1529 | 1478 | 2.22 | 4.10 | 0.27 | 0.45 |
| BF 2010 | 6959 | 230 | 77 | -1.37 | 1.64 | -1.29 | -1.44 | 66 | 1242 | 1147 | 2.42 | 4.34 | 0.16 | 0.33 |
| BJ 1996 | 2714 | 50 | 53 | -1.32 | 1.73 | -1.17 | -1.47 | 94.4 | 710 | 700 | 2.66 | 4.39 | 0.20 | 0.35 |
| BJ 2001 | 4692 | 171 | 59 | -1.51 | 1.65 | -1.44 | -1.57 | 86.5 | 743 | 792 | 2.39 | 4.29 | 0.20 | 0.35 |
| BJ 2006 | 14027 | 595 | 753 | -1.71 | 1.86 | -1.59 | -1.84 | 76.3 | 2090 | 2054 | 2.24 | 4.03 | 0.25 | 0.38 |
| BJ 2012 | 12497 | 1057 | 1267 | -1.77 | 2.47 | -1.69 | -1.85 | 70.4 | 1748 | 1768 | 2.14 | 4.01 | 0.34 | 0.43 |
| BJ 2017 | 12558 | 468 | 52 | -1.47 | 1.34 | -1.37 | -1.56 | 63.5 | 1958 | 1967 | 2.45 | 4.30 | 0.12 | 0.26 |
| BO 1994 | 3188 | 170 | 45 | -1.34 | 1.64 | -1.28 | -1.40 | 74.6 | 760 | 710 | 2.56 | 4.58 | 0.18 | 0.32 |
| BO 1998 | 6552 | 134 | 115 | -1.47 | 1.54 | -1.42 | -1.51 | 63.8 | 945 | 947 | 2.42 | 4.46 | 0.19 | 0.32 |
| BO 2003 | 9703 | 369 | 95 | -1.41 | 1.44 | -1.37 | -1.44 | 50.7 | 1539 | 1438 | 2.47 | 4.53 | 0.15 | 0.26 |
| BO 2008 | 8083 | 267 | 40 | -1.21 | 1.33 | -1.16 | -1.26 | 39.6 | 1379 | 1319 | 2.68 | 4.72 | 0.10 | 0.19 |
| BR 1996 | 4556 | 374 | 34 | -0.61 | 1.44 | -0.54 | -0.68 | 38.6 | 667 | 645 | 2.68 | 4.73 | 0.05 | 0.21 |
| BT 2007 | 2674 | 31 | 109 | -1.23 | 1.72 | -1.14 | -1.32 | 41.8 | NA | NA | NA | NA | NA | NA |
| BU 2010 | 3615 | 119 | 27 | -2.10 | 1.45 | -1.97 | -2.23 | 59.2 | 594 | 584 | 2.38 | 4.16 | 0.29 | 0.32 |
| BU 2016 | 6188 | 125 | 15 | -2.13 | 1.28 | -2.05 | -2.21 | 44.1 | 1037 | 1014 | 2.30 | 4.18 | 0.24 | 0.27 |
| CD 2007 | 3955 | 284 | 286 | -1.60 | 2.08 | -1.51 | -1.69 | 91.6 | 528 | 533 | 2.32 | 4.13 | 0.25 | 0.38 |
| CD 2013 | 8858 | 459 | 204 | -1.64 | 1.86 | -1.54 | -1.75 | 78.3 | 1308 | 1238 | 2.29 | 4.07 | 0.23 | 0.37 |
| CF 1994 | 2485 | 32 | 64 | -1.53 | 1.71 | -1.38 | -1.68 | 114.9 | 609 | 597 | 2.43 | 4.18 | 0.26 | 0.42 |
| CG 2005 | 4357 | 294 | 71 | -1.05 | 1.87 | -0.96 | -1.13 | 58.1 | 745 | 677 | 2.71 | 4.65 | 0.17 | 0.30 |
| CG 2011 | 4790 | 259 | 28 | -1.14 | 1.51 | -1.07 | -1.21 | 42 | 777 | 738 | 2.61 | 4.57 | 0.12 | 0.27 |
| CI 1994 | 3525 | 4 | 50 | -1.26 | 1.60 | -1.18 | -1.35 | 104.4 | 944 | 949 | 2.37 | 4.37 | 0.18 | 0.36 |
| CI 1998 | 1610 | 12 | 30 | -1.21 | 1.58 | -1.18 | -1.25 | 101.9 | 299 | 305 | 2.37 | 4.47 | 0.14 | 0.29 |
| CI 2012 | 3617 | 317 | 62 | -1.25 | 1.57 | -1.17 | -1.34 | 73.1 | 567 | 587 | 2.38 | 4.38 | 0.13 | 0.31 |
| CM 1991 | 2880 | 188 | 26 | -1.40 | 1.61 | -1.34 | -1.46 | 86.9 | 427 | 422 | 2.49 | 4.50 | 0.19 | 0.31 |
| CM 1998 | 1971 | 94 | 34 | -1.29 | 1.79 | -1.19 | -1.38 | 93.6 | 481 | 465 | 2.63 | 4.59 | 0.20 | 0.35 |
| CM 2004 | 3695 | 350 | 79 | -1.38 | 1.74 | -1.32 | -1.43 | 83.3 | 564 | 544 | 2.51 | 4.53 | 0.20 | 0.33 |
| CM 2011 | 5586 | 396 | 76 | -1.24 | 1.71 | -1.18 | -1.29 | 66.8 | 818 | 844 | 2.65 | 4.67 | 0.17 | 0.29 |
| CO 1995 | 4663 | 101 | 10 | -1.04 | 1.21 | -0.98 | -1.10 | 24.8 | 801 | 753 | 2.67 | 4.67 | 0.06 | 0.16 |
| CO 2000 | 4341 | 115 | 9 | -1.01 | 1.17 | -0.98 | -1.04 | 21.2 | 706 | 708 | 2.67 | 4.73 | 0.05 | 0.17 |
| CO 2005 | 13216 | 737 | 18 | -0.88 | 1.18 | -0.84 | -0.92 | 18.4 | 2164 | 2113 | 2.81 | 4.86 | 0.04 | 0.13 |
| CO 2010 | 16678 | 637 | 11 | -0.84 | 1.14 | -0.82 | -0.86 | 15.9 | 2888 | 2816 | 2.83 | 4.91 | 0.03 | 0.12 |
| DR 1991 | 3484 | 202 | 30 | -1.13 | 1.39 | -1.05 | -1.20 | 44.8 | 530 | 472 | 2.30 | 4.27 | 0.10 | 0.30 |
| DR 1996 | 4042 | 219 | 40 | -0.79 | 1.34 | -0.70 | -0.88 | 37.4 | 613 | 599 | 2.65 | 4.59 | 0.06 | 0.22 |
| DR 2002 | 10366 | 920 | 56 | -0.56 | 1.38 | -0.49 | -0.62 | 31.7 | 1517 | 1498 | 2.86 | 4.85 | 0.04 | 0.15 |
| DR 2007 | 11219 | 929 | 84 | -0.55 | 1.37 | -0.50 | -0.60 | 29.3 | 1703 | 1616 | 2.85 | 4.86 | 0.04 | 0.15 |
| DR 2013 | 4367 | 340 | 19 | -0.38 | 1.25 | -0.33 | -0.44 | 27.2 | 688 | 684 | 3.02 | 5.03 | 0.03 | 0.11 |
| EG 1992 | 7889 | 183 | 121 | -1.13 | 1.78 | -1.12 | -1.14 | 57.6 | 1248 | 1122 | 2.63 | 4.72 | 0.16 | 0.29 |
| EG 1995 | 10938 | 88 | 217 | -1.39 | 1.77 | -1.31 | -1.46 | 49.5 | 1756 | 1577 | 2.44 | 4.40 | 0.22 | 0.36 |
| EG 2000 | 10833 | 114 | 169 | -1.06 | 1.59 | -0.99 | -1.13 | 37.3 | 1771 | 1619 | 2.77 | 4.73 | 0.12 | 0.20 |
| EG 2003 | 6305 | 57 | 78 | -1.13 | 1.41 | -1.05 | -1.21 | 32.2 | 1067 | 926 | 2.70 | 4.65 | 0.10 | 0.17 |
| EG 2005 | 13262 | 86 | 379 | -1.05 | 1.91 | -0.97 | -1.12 | 29.6 | 2210 | 2100 | 2.78 | 4.73 | 0.16 | 0.26 |
| EG 2008 | 10568 | 87 | 462 | -1.05 | 2.05 | -0.97 | -1.14 | 26.2 | 1706 | 1630 | 2.78 | 4.72 | 0.20 | 0.31 |
| EG 2014 | 15361 | 146 | 637 | -0.40 | 2.16 | -0.33 | -0.46 | 20.9 | 2600 | 2380 | 3.42 | 5.40 | 0.11 | 0.17 |
| ET 2000 | 9487 | 405 | 203 | -2.07 | 1.71 | -2.02 | -2.11 | 88.2 | 1556 | 1506 | 1.87 | 3.90 | 0.34 | 0.48 |
| ET 2005 | 4413 | 208 | 241 | -1.75 | 1.96 | -1.67 | -1.83 | 70.7 | 665 | 602 | 2.22 | 4.19 | 0.31 | 0.43 |
| ET 2011 | 10571 | 681 | 176 | -1.61 | 1.79 | -1.55 | -1.66 | 52.7 | 1548 | 1477 | 2.33 | 4.36 | 0.26 | 0.39 |
| ET 2016 | 9701 | 635 | 123 | -1.37 | 1.75 | -1.32 | -1.42 | 42.5 | 1489 | 1433 | 2.57 | 4.60 | 0.20 | 0.32 |
| GA 2000 | 3630 | 58 | 26 | -1.19 | 1.61 | -1.10 | -1.28 | 54.2 | 582 | 579 | 2.66 | 4.57 | 0.14 | 0.28 |
| GA 2012 | 3894 | 403 | 71 | -1.00 | 1.60 | -0.97 | -1.04 | 40.4 | 582 | 558 | 2.79 | 4.81 | 0.11 | 0.22 |
| GH 1993 | 2056 | 89 | 34 | -1.30 | 1.66 | -1.21 | -1.38 | 73.5 | 554 | 514 | 2.44 | 4.35 | 0.19 | 0.39 |
| GH 1998 | 2894 | 36 | 62 | -1.36 | 1.64 | -1.28 | -1.43 | 67.6 | 509 | 511 | 2.36 | 4.30 | 0.15 | 0.31 |
| GH 2003 | 3372 | 171 | 44 | -1.44 | 1.59 | -1.32 | -1.56 | 58.9 | 589 | 552 | 2.33 | 4.17 | 0.16 | 0.35 |
| GH 2008 | 2730 | 191 | 100 | -1.07 | 1.71 | -1.03 | -1.12 | 52.6 | 417 | 431 | 2.62 | 4.61 | 0.10 | 0.26 |
| GH 2014 | 2868 | 129 | 4 | -0.97 | 1.31 | -0.91 | -1.03 | 40.7 | 505 | 468 | 2.74 | 4.70 | 0.05 | 0.18 |
| GM 2013 | 3720 | 346 | 175 | -1.04 | 1.66 | -0.97 | -1.11 | 45.4 | 525 | 498 | 2.59 | 4.55 | 0.11 | 0.29 |
| GN 1999 | 4743 | 74 | 178 | -1.24 | 1.86 | -1.17 | -1.29 | 104.6 | 809 | 755 | 2.34 | 4.33 | 0.18 | 0.36 |
| GN 2005 | 2922 | 175 | 52 | -1.42 | 1.86 | -1.28 | -1.56 | 82 | 478 | 463 | 2.22 | 4.06 | 0.21 | 0.44 |
| GN 2012 | 3387 | 166 | 46 | -1.06 | 1.88 | -0.97 | -1.15 | 65.4 | 595 | 539 | 2.54 | 4.47 | 0.14 | 0.31 |
| GU 1995 | 9136 | 343 | 136 | -2.36 | 1.49 | -2.29 | -2.42 | 49.2 | 1251 | 1279 | 2.25 | 4.18 | 0.42 | 0.39 |
| GU 1999 | 4453 | 418 | 64 | -2.26 | 1.46 | -2.17 | -2.35 | 42.3 | 593 | 555 | 2.37 | 4.26 | 0.37 | 0.33 |
| GU 2015 | 11996 | 207 | 18 | -1.89 | 1.20 | -1.87 | -1.91 | 24.7 | 2066 | 1983 | 2.67 | 4.70 | 0.18 | 0.16 |
| GY 2009 | 1892 | 166 | 87 | -1.03 | 1.55 | -0.96 | -1.09 | 31.2 | 291 | 301 | 2.75 | 4.84 | 0.08 | 0.17 |
| HN 2006 | 10369 | 1027 | 58 | -1.50 | 1.34 | -1.44 | -1.56 | 23.7 | 1656 | 1606 | 2.37 | 4.40 | 0.13 | 0.25 |
| HN 2012 | 10453 | 438 | 13 | -1.22 | 1.24 | -1.17 | -1.27 | 18.6 | 1886 | 1764 | 2.65 | 4.68 | 0.08 | 0.16 |
| HT 1994 | 2938 | 52 | 63 | -1.52 | 1.69 | -1.50 | -1.54 | 89.9 | 448 | 433 | 1.96 | 4.01 | 0.22 | 0.43 |
| HT 2000 | 6034 | 405 | 37 | -1.23 | 1.48 | -1.12 | -1.34 | 74.5 | 911 | 922 | 2.33 | 4.20 | 0.11 | 0.30 |
| HT 2006 | 2773 | 176 | 17 | -1.29 | 1.46 | -1.21 | -1.36 | 64.6 | 404 | 454 | 2.24 | 4.19 | 0.10 | 0.32 |
| HT 2012 | 4438 | 393 | 18 | -1.01 | 1.44 | -0.96 | -1.07 | 58.8 | 725 | 683 | 2.49 | 4.48 | 0.09 | 0.25 |
| HT 2017 | 6104 | 452 | 29 | -0.94 | 1.45 | -0.88 | -1.00 | 53.9 | 1001 | 989 | 2.57 | 4.54 | 0.08 | 0.24 |
| IA 1993 | 33000 | 3858 | 1369 | -2.07 | 1.89 | -2.02 | -2.12 | 82.1 | 5772 | 5404 | 2.23 | 4.60 | 0.35 | 0.35 |
| IA 1999 | 30435 | 3116 | 1411 | -1.89 | 1.87 | -1.86 | -1.92 | 69 | 4709 | 4341 | 2.40 | 4.79 | 0.43 | 0.43 |
| IA 2006 | 46686 | 2939 | 1299 | -1.67 | 1.75 | -1.64 | -1.69 | 53.7 | 7563 | 6511 | 2.62 | 5.03 | 0.24 | 0.24 |
| IA 2015 | 247577 | 10463 | 4460 | -1.42 | 1.77 | -1.38 | -1.46 | 35.3 | 41384 | 36372 | 2.87 | 5.25 | 0.18 | 0.18 |
| JO 1990 | 7454 | 565 | 86 | -0.93 | 1.47 | -0.91 | -0.95 | 29.8 | 899 | 841 | 2.62 | 4.72 | 0.08 | 0.22 |
| JO 1997 | 6097 | 405 | 44 | -0.71 | 1.24 | -0.69 | -0.73 | 25 | 811 | 740 | 2.84 | 4.93 | 0.03 | 0.12 |
| JO 2002 | 5607 | 671 | 20 | -0.70 | 1.25 | -0.72 | -0.68 | 22.2 | 759 | 707 | 2.81 | 4.99 | 0.03 | 0.12 |
| JO 2007 | 5151 | 374 | 192 | -0.56 | 1.84 | -0.54 | -0.58 | 19.5 | 688 | 664 | 2.99 | 5.08 | 0.09 | 0.19 |
| JO 2009 | 4686 | 257 | 31 | -0.53 | 1.27 | -0.50 | -0.56 | 18.5 | NA | NA | 3.03 | 5.10 | NA | NA |
| JO 2012 | 6644 | 286 | 49 | -0.47 | 1.22 | -0.45 | -0.49 | 17 | 1051 | 927 | 3.08 | 5.17 | 0.02 | 0.08 |
| KE 1993 | 5330 | 231 | 102 | -1.60 | 1.59 | -1.48 | -1.71 | 69.6 | 789 | 776 | 2.34 | 4.14 | 0.21 | 0.38 |
| KE 1998 | 3211 | 98 | 101 | -1.39 | 1.85 | -1.26 | -1.52 | 67.7 | 791 | 795 | 2.56 | 4.33 | 0.23 | 0.37 |
| KE 2009 | 5660 | 316 | 137 | -1.33 | 1.74 | -1.25 | -1.39 | 41.4 | 837 | 821 | 2.57 | 4.46 | 0.18 | 0.33 |
| KE 2014 | 19820 | 875 | 124 | -1.17 | 1.45 | -1.06 | -1.28 | 36.3 | 3292 | 3136 | 2.76 | 4.57 | 0.11 | 0.23 |

| | | | | | | | | | | | | | |
|---|---|---|---|---|---|---|---|---|---|---|---|---|---|
| KH 2000 | 3858 | 86 | 98 | -1.87 | 1.86 | -1.85 | -1.88 | 79.6 | 536 | 559 | 2.13 | 4.55 | 0.28 | 0.36 |
| KH 2005 | 3796 | 108 | 67 | -1.85 | 1.40 | -1.79 | -1.92 | 53.3 | 631 | 619 | 2.18 | 4.51 | 0.19 | 0.25 |
| KH 2010 | 3910 | 102 | 61 | -1.63 | 1.47 | -1.63 | -1.63 | 37.8 | 698 | 677 | 2.34 | 4.79 | 0.16 | 0.20 |
| KH 2014 | 4554 | 127 | 58 | -1.39 | 1.44 | -1.35 | -1.43 | 28.9 | 844 | 789 | 2.62 | 5.00 | 0.11 | 0.16 |
| KK 1995 | 774 | 26 | 7 | -0.79 | 1.52 | -0.63 | -0.97 | 44.8 | 190 | 216 | 3.02 | 4.85 | 0.08 | 0.21 |
| KK 1999 | 598 | 19 | 5 | -0.74 | 1.42 | -0.76 | -0.72 | 39.1 | 101 | 108 | 2.90 | 5.11 | 0.06 | 0.12 |
| KM 1996 | 1035 | 33 | 16 | -1.58 | 1.69 | -1.51 | -1.65 | 75.5 | 245 | 200 | 2.33 | 4.20 | 0.24 | 0.39 |
| KM 2012 | 2956 | 241 | 160 | -1.07 | 2.05 | -0.96 | -1.19 | 59.5 | 394 | 370 | 2.89 | 4.66 | 0.18 | 0.27 |
| KY 1997 | 1022 | 32 | 7 | -1.28 | 1.57 | -1.15 | -1.39 | 47.5 | 293 | 250 | 2.80 | 4.71 | 0.12 | 0.22 |
| KY 2012 | 4229 | 161 | 14 | -0.82 | 1.48 | -0.75 | -0.89 | 23.2 | 664 | 618 | 3.20 | 5.21 | 0.09 | 0.13 |
| LB 2007 | 5178 | 584 | 151 | -1.48 | 1.83 | -1.38 | -1.57 | 80.2 | 760 | 746 | 2.25 | 4.01 | 0.21 | 0.40 |
| LB 2013 | 3583 | 321 | 49 | -1.28 | 1.65 | -1.18 | -1.37 | 63.2 | 570 | 506 | 2.45 | 4.21 | 0.13 | 0.29 |
| LS 2004 | 1642 | 173 | 61 | -1.77 | 1.62 | -1.72 | -1.82 | 85.7 | 257 | 246 | 2.32 | 4.30 | 0.25 | 0.35 |
| LS 2009 | 1848 | 173 | 32 | -1.54 | 1.59 | -1.42 | -1.67 | 79.2 | 281 | 325 | 2.62 | 4.45 | 0.16 | 0.24 |
| LS 2014 | 1497 | 147 | 18 | -1.50 | 1.36 | -1.40 | -1.61 | 70.9 | 250 | 272 | 2.63 | 4.51 | 0.13 | 0.23 |
| MA 1992 | 4744 | 45 | 96 | -1.28 | 1.56 | -1.25 | -1.30 | 58.1 | 753 | 721 | 2.10 | 4.27 | 0.15 | 0.38 |
| MA 2003 | 5760 | 79 | 111 | -0.85 | 1.81 | -0.78 | -0.92 | 37.1 | 980 | 946 | 2.57 | 4.65 | 0.14 | 0.28 |
| MB 2005 | 1477 | 95 | 39 | -0.15 | 1.70 | -0.14 | -0.17 | 16.8 | 276 | 265 | 3.08 | 5.14 | 0.04 | 0.14 |
| MD 1992 | 4356 | 114 | 50 | -2.29 | 1.45 | -2.16 | -2.41 | 92.4 | 676 | 628 | 2.02 | 3.80 | 0.38 | 0.47 |
| MD 1997 | 3189 | 93 | 41 | -2.04 | 1.57 | -1.86 | -2.22 | 79 | 767 | 766 | 2.31 | 3.98 | 0.36 | 0.45 |
| MD 2004 | 4988 | 247 | 148 | -1.87 | 1.79 | -1.77 | -1.99 | 56.6 | 717 | 740 | 2.41 | 4.22 | 0.30 | 0.38 |
| MD 2009 | 5837 | 299 | 299 | -1.71 | 2.02 | -1.64 | -1.78 | 45 | 849 | 830 | 2.54 | 4.43 | 0.28 | 0.34 |
| ML 1996 | 5114 | 107 | 96 | -1.36 | 1.85 | -1.29 | -1.44 | 123.1 | 1221 | 1212 | 2.52 | 4.43 | 0.27 | 0.42 |
| ML 2001 | 10555 | 528 | 335 | -1.60 | 1.89 | -1.54 | -1.67 | 111.7 | 1567 | 1526 | 2.26 | 4.21 | 0.29 | 0.44 |
| ML 2006 | 12201 | 558 | 301 | -1.40 | 2.01 | -1.32 | -1.47 | 90.7 | 1969 | 1875 | 2.48 | 4.41 | 0.25 | 0.39 |
| ML 2012 | 5007 | 411 | 148 | -1.39 | 1.95 | -1.34 | -1.44 | 74.7 | 734 | 778 | 2.47 | 4.44 | 0.23 | 0.35 |

| | | | | | | | | | | | | | |
|---|---|---|---|---|---|---|---|---|---|---|---|---|---|
| MM 2016 | 4504 | 273 | 15 | -1.37 | 1.32 | -1.32 | -1.41 | 39.9 | 780 | 711 | 2.86 | 5.17 | 0.10 | 0.11 |
| MR 2000 | 4093 | 76 | 144 | -1.39 | 1.98 | -1.34 | -1.44 | 69.7 | 625 | 577 | 2.23 | 4.13 | 0.25 | 0.46 |
| MV 2009 | 2917 | 460 | 49 | -0.90 | 1.47 | -0.88 | -0.92 | 12.4 | 426 | 457 | 3.16 | 5.65 | 0.06 | 0.08 |
| MV 2017 | 2819 | 389 | 55 | -0.85 | 1.28 | -0.85 | -0.86 | 6.8 | 424 | 437 | 3.19 | 5.71 | 0.06 | 0.07 |
| MW 1992 | 3499 | 134 | 58 | -2.03 | 1.57 | -1.91 | -2.14 | 129.1 | 548 | 505 | 2.13 | 3.93 | 0.30 | 0.44 |
| MW 2000 | 10294 | 521 | 364 | -1.96 | 1.77 | -1.87 | -2.06 | 101.1 | 1648 | 1698 | 2.17 | 4.02 | 0.32 | 0.44 |
| MW 2004 | 9565 | 848 | 401 | -1.96 | 1.78 | -1.85 | -2.07 | 74 | 1470 | 1455 | 2.19 | 4.00 | 0.31 | 0.43 |
| MW 2010 | 6239 | 1378 | 150 | -1.78 | 1.66 | -1.66 | -1.90 | 56.2 | 890 | 847 | 2.38 | 4.17 | 0.23 | 0.36 |
| MW 2015 | 5564 | 314 | 84 | -1.51 | 1.39 | -1.46 | -1.57 | 42 | 922 | 986 | 2.58 | 4.51 | 0.11 | 0.21 |
| MZ 1997 | 3621 | 39 | 80 | -1.69 | 1.87 | -1.54 | -1.84 | 131.7 | 893 | 942 | 2.43 | 4.18 | 0.31 | 0.43 |
| MZ 2003 | 8678 | 384 | 148 | -1.84 | 1.52 | -1.75 | -1.93 | 97.4 | 1407 | 1374 | 2.23 | 4.09 | 0.26 | 0.37 |
| MZ 2011 | 10241 | 507 | 220 | -1.60 | 1.65 | -1.51 | -1.69 | 66.9 | 1620 | 1718 | 2.47 | 4.34 | 0.20 | 0.31 |
| NC 1998 | 7468 | 285 | 173 | -1.39 | 1.54 | -1.33 | -1.46 | 36.1 | 1062 | 1074 | 2.22 | 4.26 | 0.16 | 0.33 |
| NC 2001 | 6477 | 362 | 99 | -1.18 | 1.49 | -1.15 | -1.20 | 29.9 | 1054 | 1018 | 2.40 | 4.51 | 0.08 | 0.23 |
| NG 1990 | 6416 | 267 | 197 | -1.83 | 1.84 | -1.76 | -1.89 | 125.6 | 956 | 927 | 2.30 | 4.19 | 0.30 | 0.40 |
| NG 2003 | 5008 | 206 | 243 | -1.60 | 1.97 | -1.49 | -1.70 | 102.5 | 728 | 724 | 2.57 | 4.38 | 0.30 | 0.38 |
| NG 2008 | 24759 | 1511 | 2593 | -1.49 | 2.22 | -1.39 | -1.59 | 86.8 | 3185 | 3221 | 2.67 | 4.49 | 0.31 | 0.39 |
| NG 2013 | 28312 | 1458 | 1377 | -1.29 | 2.07 | -1.21 | -1.37 | 73.3 | 4193 | 3965 | 2.85 | 4.71 | 0.25 | 0.32 |
| NI 1992 | 5226 | 339 | 84 | -1.70 | 1.80 | -1.62 | -1.78 | 126.7 | 830 | 696 | 2.29 | 4.22 | 0.30 | 0.43 |
| NI 1998 | 4119 | 84 | 53 | -1.76 | 1.72 | -1.66 | -1.86 | 103 | 1038 | 953 | 2.25 | 4.13 | 0.33 | 0.47 |
| NI 2006 | 4070 | 205 | 102 | -1.92 | 1.84 | -1.83 | -2.02 | 74.5 | 638 | 644 | 2.08 | 3.97 | 0.35 | 0.49 |
| NI 2012 | 5600 | 451 | 218 | -1.63 | 1.78 | -1.55 | -1.70 | 57.6 | 752 | 761 | 2.36 | 4.29 | 0.26 | 0.38 |
| NM 1992 | 2892 | 88 | 72 | -1.45 | 1.60 | -1.33 | -1.58 | 47.3 | 464 | 488 | 2.45 | 4.28 | 0.18 | 0.33 |
| NM 2000 | 3713 | 667 | 49 | -1.13 | 1.55 | -1.04 | -1.22 | 47.9 | 523 | 512 | 2.74 | 4.64 | 0.12 | 0.25 |
| NM 2007 | 4755 | 908 | 85 | -1.24 | 1.50 | -1.18 | -1.30 | 39.5 | 680 | 652 | 2.60 | 4.56 | 0.14 | 0.28 |
| NM 2013 | 2392 | 510 | 45 | -0.94 | 1.52 | -0.91 | -0.97 | 37 | 325 | 336 | 2.88 | 4.89 | 0.10 | 0.21 |
| NP 1996 | 4059 | 246 | 48 | -2.18 | 1.42 | -2.12 | -2.23 | 73.3 | 1039 | 968 | 1.97 | 4.34 | 0.37 | 0.42 |
| NP 2001 | 6361 | 108 | 38 | -2.18 | 1.34 | -2.17 | -2.19 | 57.3 | 1024 | 1058 | 1.91 | 4.37 | 0.29 | 0.34 |
| NP 2006 | 5405 | 126 | 16 | -1.96 | 1.34 | -1.97 | -1.96 | 45.2 | 930 | 914 | 2.12 | 4.61 | 0.22 | 0.26 |
| NP 2011 | 2424 | 64 | 14 | -1.71 | 1.39 | -1.67 | -1.74 | 35.8 | 425 | 373 | 2.42 | 4.83 | 0.17 | 0.21 |
| NP 2016 | 2444 | 65 | 7 | -1.52 | 1.35 | -1.50 | -1.54 | 28.8 | 488 | 392 | 2.58 | 5.02 | 0.12 | 0.15 |
| PE 1991 | 8248 | 373 | 73 | -1.59 | 1.45 | -1.52 | -1.65 | 54.3 | 1215 | 1160 | 2.52 | 4.50 | 0.18 | 0.27 |
| PE 1996 | 15699 | 435 | 156 | -1.45 | 1.52 | -1.39 | -1.51 | 40.2 | 2505 | 2420 | 2.66 | 4.64 | 0.16 | 0.24 |
| PE 2000 | 12382 | 587 | 73 | -1.47 | 1.42 | -1.46 | -1.48 | 29.6 | 2033 | 2000 | 2.59 | 4.68 | 0.13 | 0.21 |
| PE 2006 | 2522 | 201 | 6 | -1.43 | 1.26 | -1.35 | -1.51 | 19.6 | 424 | 374 | 2.69 | 4.65 | 0.11 | 0.18 |
| PE 2008 | 8833 | 659 | 11 | -1.34 | 1.21 | -1.30 | -1.37 | 17.5 | 1464 | 1506 | 2.75 | 4.78 | 0.09 | 0.15 |
| PE 2009 | 9872 | 465 | 5 | -1.30 | 1.19 | -1.27 | -1.33 | 16.6 | 1664 | 1603 | 2.78 | 4.82 | 0.09 | 0.14 |
| PE 2010 | 9032 | 228 | 7 | -1.28 | 1.13 | -1.26 | -1.31 | 15.8 | 1642 | 1587 | 2.79 | 4.85 | 0.07 | 0.12 |
| PE 2011 | 8924 | 173 | 7 | -1.23 | 1.09 | -1.20 | -1.25 | 15.1 | 1671 | 1609 | 2.84 | 4.91 | 0.05 | 0.11 |
| PE 2012 | 9419 | 191 | 3 | -1.15 | 1.07 | -1.13 | -1.18 | 14.4 | 1741 | 1669 | 2.91 | 4.98 | 0.05 | 0.10 |
| PK 1991 | 5508 | 821 | 270 | -2.22 | 1.87 | -2.18 | -2.25 | 104.6 | 663 | 666 | 2.12 | 4.54 | 0.38 | 0.37 |
| PK 2012 | 3813 | 150 | 420 | -1.79 | 1.98 | -1.74 | -1.84 | 69.7 | 488 | 440 | 2.56 | 4.96 | 0.30 | 0.29 |
| PK 2018 | 4413 | 185 | 85 | -1.55 | 1.69 | -1.54 | -1.56 | 61.2 | 528 | 520 | 2.76 | 5.23 | 0.21 | 0.20 |
| PY 1990 | 3921 | 239 | 21 | -1.00 | 1.30 | -0.93 | -1.06 | 36.3 | 538 | 504 | 2.58 | 4.59 | 0.08 | 0.23 |
| RW 1992 | 4602 | 183 | 56 | -2.07 | 1.52 | -1.99 | -2.16 | 100.3 | 762 | 742 | 2.11 | 3.99 | 0.30 | 0.41 |
| RW 2000 | 6699 | 303 | 151 | -1.69 | 1.80 | -1.59 | -1.79 | 107.7 | 908 | 913 | 2.51 | 4.36 | 0.27 | 0.37 |
| RW 2005 | 3915 | 124 | 89 | -1.91 | 1.61 | -1.84 | -1.99 | 69.1 | 609 | 613 | 2.26 | 4.17 | 0.28 | 0.39 |
| RW 2010 | 4239 | 117 | 18 | -1.74 | 1.41 | -1.62 | -1.86 | 44.5 | 728 | 701 | 2.48 | 4.29 | 0.21 | 0.29 |
| RW 2015 | 7498 | 296 | 76 | -1.55 | 1.40 | -1.42 | -1.68 | 31.4 | 1338 | 1242 | 2.68 | 4.48 | 0.16 | 0.23 |
| SL 2008 | 2527 | 242 | 102 | -1.25 | 2.20 | -1.13 | -1.37 | 116.9 | 379 | 398 | 2.47 | 4.33 | 0.23 | 0.37 |
| SL 2013 | 5231 | 483 | 425 | -1.31 | 2.10 | -1.28 | -1.33 | 95.8 | 705 | 783 | 2.32 | 4.36 | 0.21 | 0.39 |
| SN 1993 | 4818 | 123 | 97 | -1.38 | 1.64 | -1.34 | -1.43 | 70.6 | 751 | 722 | 2.08 | 4.14 | 0.19 | 0.39 |
| SN 2005 | 3276 | 335 | 38 | -0.92 | 1.50 | -0.86 | -0.97 | 54 | 515 | 448 | 2.56 | 4.60 | 0.08 | 0.25 |

| | | | | | | | | | | | | | |
|---|---|---|---|---|---|---|---|---|---|---|---|---|---|
| SN 2010 | 4459 | 531 | 124 | -1.25 | 1.65 | -1.17 | -1.32 | 42.7 | 702 | 593 | 2.25 | 4.25 | 0.16 | 0.37 |
| SN 2012 | 6442 | 371 | 83 | -0.97 | 1.38 | -0.88 | -1.07 | 39.3 | 996 | 1038 | 2.54 | 4.50 | 0.10 | 0.26 |
| SN 2014 | 6455 | 345 | 46 | -1.06 | 1.30 | -1.00 | -1.11 | 36.3 | 1084 | 1056 | 2.42 | 4.46 | 0.08 | 0.26 |
| SN 2015 | 6560 | 325 | 61 | -1.10 | 1.26 | -1.06 | -1.14 | 34.9 | 1081 | 1081 | 2.35 | 4.43 | 0.07 | 0.25 |
| SN 2016 | 6362 | 297 | 32 | -1.01 | 1.26 | -0.92 | -1.08 | 33.8 | 1023 | 1018 | 2.50 | 4.49 | 0.06 | 0.22 |
| SN 2017 | 11459 | 627 | 71 | -0.97 | 1.29 | -0.92 | -1.02 | 32.7 | 1895 | 1821 | 2.50 | 4.55 | 0.06 | 0.24 |
| ST 2008 | 1784 | 72 | 118 | -1.06 | 1.92 | -1.00 | -1.13 | 36.9 | 299 | 305 | 2.52 | 4.38 | 0.13 | 0.31 |
| SZ 2006 | 2489 | 379 | 40 | -1.19 | 1.45 | -1.08 | -1.29 | 72.6 | 383 | 343 | 2.84 | 4.75 | 0.13 | 0.22 |
| TD 1997 | 6060 | 203 | 105 | -1.61 | 1.91 | -1.56 | -1.66 | 104.8 | 881 | 905 | 2.38 | 4.41 | 0.30 | 0.40 |
| TD 2004 | 4758 | 103 | 109 | -1.54 | 2.10 | -1.48 | -1.60 | 95.1 | 719 | 717 | 2.46 | 4.47 | 0.34 | 0.44 |
| TD 2015 | 10827 | 398 | 241 | -1.60 | 1.99 | -1.55 | -1.65 | 76.7 | 1479 | 1562 | 2.39 | 4.42 | 0.31 | 0.42 |
| TG 1998 | 3827 | 39 | 64 | -1.23 | 1.68 | -1.12 | -1.35 | 79.6 | 986 | 1001 | 2.36 | 4.17 | 0.19 | 0.43 |
| TG 2014 | 3351 | 119 | 19 | -1.27 | 1.41 | -1.23 | -1.30 | 53.3 | 584 | 587 | 2.25 | 4.22 | 0.09 | 0.29 |
| TJ 2012 | 4809 | 40 | 127 | -1.07 | 1.68 | -1.05 | -1.08 | 34.3 | 819 | 739 | 2.77 | 4.95 | 0.11 | 0.20 |
| TJ 2017 | 5999 | 85 | 33 | -0.82 | 1.49 | -0.82 | -0.82 | 29.4 | 982 | 960 | 3.00 | 5.21 | 0.06 | 0.12 |
| TL 2009 | 9148 | 960 | 359 | -2.02 | 1.98 | -1.94 | -2.10 | 54.1 | 1202 | 1175 | 2.61 | 4.70 | 0.37 | 0.32 |
| TL 2016 | 6895 | 679 | 297 | -1.53 | 2.15 | -1.45 | -1.62 | 42.2 | 965 | 918 | 3.10 | 5.18 | 0.28 | 0.24 |
| TR 1993 | 3400 | 207 | 28 | -0.95 | 1.50 | -0.96 | -0.95 | 47.6 | 539 | 452 | 2.44 | 4.60 | 0.07 | 0.23 |
| TR 1998 | 3268 | 425 | 16 | -0.85 | 1.49 | -0.86 | -0.85 | 36 | 509 | 411 | 2.54 | 4.71 | 0.07 | 0.20 |
| TR 2004 | 4303 | 235 | 10 | -0.70 | 1.48 | -0.74 | -0.65 | 24.6 | NA | NA | 2.66 | 4.90 | NA | NA |
| TR 2008 | 3570 | 769 | 30 | -0.65 | 1.45 | -0.67 | -0.63 | 18.8 | NA | NA | 2.72 | 4.93 | NA | NA |
| TR 2013 | 1733 | 260 | 89 | -0.19 | 1.61 | -0.09 | -0.28 | 13.3 | NA | NA | 3.30 | 5.27 | NA | NA |
| TZ 1991 | 6800 | 41 | 176 | -1.97 | 1.55 | -1.89 | -2.06 | 99.3 | 1133 | 1163 | 2.21 | 4.13 | 0.29 | 0.38 |
| TZ 1996 | 5678 | 58 | 180 | -1.96 | 1.56 | -1.86 | -2.05 | 92.8 | 949 | 951 | 2.24 | 4.14 | 0.30 | 0.39 |
| TZ 1999 | 2805 | 217 | 32 | -1.87 | 1.39 | -1.79 | -1.94 | 83.6 | 444 | 429 | 2.31 | 4.25 | 0.24 | 0.33 |
| TZ 2004 | 7735 | 431 | 75 | -1.74 | 1.40 | -1.65 | -1.83 | 62.7 | 1247 | 1241 | 2.45 | 4.36 | 0.20 | 0.29 |
| TZ 2010 | 7376 | 419 | 95 | -1.62 | 1.49 | -1.53 | -1.70 | 48.1 | 1081 | 1145 | 2.57 | 4.49 | 0.21 | 0.29 |
| TZ 2015 | 9623 | 571 | 51 | -1.43 | 1.40 | -1.35 | -1.51 | 41 | 1623 | 1576 | 2.75 | 4.68 | 0.14 | 0.21 |
| UG 1995 | 4841 | 63 | 122 | -1.68 | 1.60 | -1.54 | -1.83 | 98 | 870 | 915 | 2.27 | 4.01 | 0.22 | 0.40 |
| UG 2000 | 6028 | 745 | 84 | -1.75 | 1.52 | -1.68 | -1.82 | 88.4 | 827 | 840 | 2.13 | 4.02 | 0.23 | 0.41 |
| UG 2006 | 2603 | 182 | 29 | -1.53 | 1.58 | -1.46 | -1.60 | 64 | 338 | 369 | 2.34 | 4.24 | 0.16 | 0.33 |
| UG 2011 | 2261 | 151 | 17 | -1.36 | 1.58 | -1.25 | -1.48 | 48.3 | 310 | 342 | 2.56 | 4.36 | 0.16 | 0.29 |
| UG 2016 | 4903 | 447 | 27 | -1.21 | 1.45 | -1.12 | -1.31 | 37 | 771 | 699 | 2.69 | 4.54 | 0.11 | 0.25 |
| UZ 1996 | 1221 | 129 | 68 | -1.28 | 2.16 | -1.04 | -1.50 | 57.1 | 283 | 285 | 2.98 | 4.54 | 0.23 | 0.30 |
| YE 1991 | 5573 | 2614 | 14 | -1.81 | 1.65 | -1.98 | -1.59 | 86.2 | 379 | 458 | 2.00 | 4.77 | 0.29 | 0.37 |
| YE 2013 | 15115 | 828 | 340 | -1.81 | 1.69 | -1.77 | -1.85 | 43.2 | 2219 | 1984 | 2.21 | 4.51 | 0.25 | 0.31 |
| ZA 2016 | 1575 | 445 | 16 | -1.15 | 1.43 | -1.13 | -1.16 | 30 | 217 | 193 | 2.82 | 5.04 | 0.12 | 0.19 |
| ZM 1992 | 5148 | 60 | 76 | -1.88 | 1.49 | -1.76 | -2.00 | 110.7 | 837 | 847 | 2.40 | 4.20 | 0.27 | 0.36 |
| ZM 1996 | 5793 | 95 | 100 | -1.95 | 1.56 | -1.88 | -2.02 | 105.5 | 961 | 1018 | 2.28 | 4.17 | 0.30 | 0.40 |
| ZM 2002 | 5877 | 234 | 132 | -2.05 | 1.66 | -1.95 | -2.14 | 85.3 | 966 | 997 | 2.21 | 4.06 | 0.33 | 0.41 |
| ZM 2007 | 5751 | 358 | 173 | -1.66 | 1.78 | -1.53 | -1.79 | 60.9 | 864 | 882 | 2.63 | 4.40 | 0.23 | 0.31 |
| ZM 2013 | 12407 | 605 | 190 | -1.55 | 1.68 | -1.46 | -1.64 | 48.3 | 2027 | 2061 | 2.70 | 4.56 | 0.22 | 0.29 |
| ZW 1994 | 2164 | 14 | 25 | -1.19 | 1.54 | -1.10 | -1.29 | 56.6 | 562 | 598 | 2.67 | 4.55 | 0.13 | 0.30 |
| ZW 1999 | 3210 | 360 | 107 | -1.18 | 1.84 | -1.08 | -1.27 | 58.9 | 518 | 538 | 2.69 | 4.57 | 0.17 | 0.32 |
| ZW 2005 | 4697 | 480 | 178 | -1.37 | 1.67 | -1.32 | -1.43 | 59 | 715 | 732 | 2.45 | 4.40 | 0.18 | 0.32 |
| ZW 2010 | 5043 | 631 | 55 | -1.34 | 1.47 | -1.28 | -1.41 | 55.1 | 811 | 777 | 2.49 | 4.43 | 0.16 | 0.33 |
| ZW 2015 | 5676 | 672 | 43 | -1.20 | 1.39 | -1.12 | -1.28 | 40.2 | 950 | 932 | 2.65 | 4.56 | 0.12 | 0.25 |

**Variables Descriptions**

- *Height-for-Age Z-scores (HFZ z-score)*:  We used the WHO SPSS anthro macros (http://www.who.int/childgrowth/software/en/) to estimate HAZ for all children in the full sample.  Consistent with Jayachandran and Pande, we follow WHO guidelines and excluded children with implausible anthropometric values of +/- 6 SD.

The explanatory variables broadly represent sources of influence on childhood growth, ranging from resource access and prenatal and postnatal care to hygiene and infectious disease exposure (Headey, Hoddinott, & Park, 2016; B. S. Jayachandran et al., 2017). We indicate whether these variables are measured at the household, sampling cluster, or country-level.

*Child characteristics*

- *Age.* We include child age as a continuous variable.  To account for known nonlinear associations between child age and HAZ, we first center age at 24 m.  This is near the age at which HAZ stops declining with age (Leroy et al 2014 and Leroy et al 2015).  Additionally, we include a spline set at 24 m.  The spline permits estimating the effect of child age on HAZ before 24 m and after 24 m independently.

- *Birth Order.* We control for birth-order effects on child height using a series of dummy variables indicating if the child was born second or third-or-greater (with 1[st] born as the reference category) (S. Jayachandran & Pande, 2013).

- *Antenatal visits.* Indicates if a child's mother had four or more antenatal visits.

- *Facility Birth.* We include a binary variable indicating if the child was born in a medical facility.

- *Vaccinations.* We also include a binary variable indicating if a child had more than 7 total vaccinations including BCG, 3 DPT shots, 3 polio shots, and 1 measles shot.  For the sensitivity analysis with the most recent birth, we also include the following additional covariates:

- *Mother Iron supplements during pregnancy:* a binary variable indicating if the mother was taking iron supplements during her pregnancy.

- *Child Iron supplements:* a binary variable indicating if the child had taken iron supplements.

- *Nutrition Variables.* Recent studies have suggested that animal source proteins in the diet may improve child growth, and may in fact account for growth differences between sub-Saharan Africa and South Asia (Baten & Blum, 2012; Grasgruber, Cacek, Kalina, & Sebera, 2014). For a subset of the DHS studies, information was collected on whether children were fed a range of foods in the previous 24 hours. These were coded into two variables—non-dairy animal source (e.g., meat, fish, poultry, innards, and animal-derived products) and dairy products or substitutes (e.g., milk, cheese, yogurt, formula).

<u>Maternal characteristics</u>

- *Education:* Mother's education was dummy coded based on four categories—no schooling, primary school, secondary school, post-secondary school—with no schooling as the reference category.
- *Mother's parity:* The number of live born children was included and top coded at 12 children.
- *Maternal literacy:* In rare surveys where this was missing, this was imputed based on stage of education achieved and years of schooling.
- *Mother's age:* Mother's age (centered at 30 y) was included as both a linear and quadratic term to capture nonlinear associations between maternal age and child height.

<u>Household characteristics</u>

- *Urban residence.*
- *Household open defecation.*
- *Absolute Wealth Estimates (AWE)—Household-level:* Using an asset-based approach, we estimated the household wealth per capita in absolute units— 2011-constant international dollars with purchasing power parity (Hruschka, Gerkey, & Hadley, 2015). This facilitates comparisons of household wealth both within a country across different survey years, as well as across survey populations. We used two measures of household wealth per capita: (1) a continuous log-transformed value for AWE and (2) a categorical variable binned into sixteen categories, each representing a 50% increase over the prior category [< $90 through > approx. $36,000].

- *Study Year.* We include a year variable indicating year since 1990, our earliest set of surveys, to capture any potential increases in HAZ over time that are not captured by our explanatory variables, and to ensure our estimates of basal HAZ are set to the lowest estimated value.

<u>*Cluster and subdistrict-level characteristics*</u>

- *Open defecation.* We control for the impact of sanitation on child height using a cluster level variable of the proportion of households in the cluster who engage in open defecation. Headey et al. reported nonlinearities in the association between cluster-level measures of open defecation and HAZ measurements (Headey et al., 2016). Specifically, they found steep negative association for proportions below 0.30, after which the association becomes relatively flat. To address these nonlinearities, we include two variables capturing cluster-level exposure to open defecation. The first is the raw continuous proportion, ranging from 0 to 1. The second is a spline of the raw proportion centered on 0.3 proportion and indicates how the effect of open-defecation changes as the proportion exceeds 0.3.
- *Exposure to infectious disease*—To account for infectious disease exposure, we include cluster-, and subdistrict-level proportions of the households with a child who experienced diarrhea within the last two week.

<u>*Subdistricts, States and Countries.*</u> Subdistricts were the 1$^{st}$ level administrative districts in a country. For some small Indian states, we combined them with neighboring larger states. Nicobar and Andaman Islands->W. Bengal; Lakshadweep ->Kerala; Chandigarh->Punjab; Dadra & Nagar -> Maharashtra; Daman -> Gujarat; Goa->Karnataka; Pondicherry->Tamil

**Table S2. Variables descriptives by region (main model).**

| | S & SE Asia<br>N=148437 | sub-Saharan Africa<br>N=224069 | Other Regions<br>N=142169 |
|---|---|---|---|
| HFA z-scores | -1.78 (1.77) | -1.79 (1.70) | -1.24 (1.57) |
| Child's age | 22.4 (6.84) | 21.9 (6.71) | 22.4 (6.82) |
| mother's parity | 2.62 (1.78) | 3.83 (2.50) | 3.13 (2.27) |
| mother's age | 27.0 (5.59) | 28.8 (6.96) | 28.6 (6.65) |
| mother literate | 59.4% | 43.9% | 79.0% |
| no schooling | 36.3% | 45.1% | 16.8% |
| primary school | 18.3% | 35.8% | 32.6% |
| secondary school | 37.3% | 17.3% | 38.3% |
| post-secondary | 8.13% | 1.79% | 12.3% |
| ln(household wealth) | 8.06 (1.24) | 6.99 (1.35) | 8.13 (1.35) |
| open defecation | 43.8% | 34.1% | 16.5% |
| % diarrhea community | 0.10 (0.11) | 0.15 (0.12) | 0.16 (0.15) |
| > 3 antenatal visits | 41.6% | 49.5% | 68.6% |
| facility birth | 57.2% | 54.2% | 68.9% |
| > 7 vaccines | 59.0% | 53.1% | 64.3% |
| urban residence | 26.2% | 28.9% | 47.9% |
| survey year | 2010 (7.65) | 2006 (7.65) | 2004 (7.42) |
| 1st born | 30.7% | 19.6% | 26.6% |
| 2nd born | 29.3% | 18.3% | 24.2% |
| > 2nd born | 40.0% | 62.1% | 49.2% |

**Table S3. Results of linear models**

| Predictors | Boys 12-35 | | Girls 12-35 | |
|---|---|---|---|---|
| | Estimates | CI | Estimates | CI |
| (Intercept) | -2.86 *** | -2.97 – -2.76 | -2.63 *** | -2.74 – -2.53 |
| ln(household wealth) | 0.14 *** | 0.14 – 0.15 | 0.14 *** | 0.13 – 0.15 |
| mother's parity | -0.05 *** | -0.05 – -0.04 | -0.05 *** | -0.05 – -0.04 |
| open defecation | -0.04 *** | -0.07 – -0.02 | -0.02 * | -0.05 – -0.00 |
| % Open Defecation < | -0.17 *** | -0.25 – -0.09 | -0.22 *** | -0.30 – -0.14 |
| % Open Defecation > | 0.19 *** | 0.09 – 0.30 | 0.21 *** | 0.11 – 0.32 |
| mother literate | 0.08 *** | 0.06 – 0.10 | 0.06 *** | 0.04 – 0.09 |
| primary school | 0.05 *** | 0.03 – 0.07 | 0.07 *** | 0.05 – 0.09 |
| secondary school | 0.22 *** | 0.19 – 0.24 | 0.22 *** | 0.19 – 0.25 |
| post-secondary | 0.38 *** | 0.35 – 0.42 | 0.40 *** | 0.36 – 0.44 |
| Mother age | 0.23 *** | 0.21 – 0.24 | 0.24 *** | 0.23 – 0.26 |
| Mother age^2 | -0.05 *** | -0.07 – -0.04 | -0.04 *** | -0.06 – -0.03 |
| Child age > 24 m | 0.00 * | 0.00 – 0.00 | -0.01 *** | -0.01 – -0.01 |
| Child age < 24 m | -0.04 *** | -0.05 – -0.04 | -0.04 *** | -0.04 – -0.03 |
| > 7 vaccines | 0.03 *** | 0.02 – 0.04 | 0.05 *** | 0.03 – 0.06 |
| facility birth | 0.13 *** | 0.11 – 0.15 | 0.13 *** | 0.11 – 0.14 |
| Delivery place | 0.06 | -0.02 – 0.14 | 0.07 | -0.01 – 0.15 |
| > 3 antenatal visits | 0.12 *** | 0.11 – 0.14 | 0.12 *** | 0.10 – 0.13 |
| Antenatal missing | 0.09 *** | 0.04 – 0.14 | 0.12 *** | 0.07 – 0.17 |
| Birth order 2 | -0.05 *** | -0.07 – -0.03 | -0.07 *** | -0.09 – -0.06 |
| Birth order > 2 | -0.10 *** | -0.12 – -0.07 | -0.13 *** | -0.16 – -0.11 |
| % diarrhea | -0.23 *** | -0.28 – -0.17 | -0.19 *** | -0.25 – -0.13 |
| % diarrhea subdistrict | -0.50 *** | -0.65 – -0.35 | -0.42 *** | -0.57 – -0.27 |
| urban residence | 0.03 *** | 0.02 – 0.05 | 0.03 *** | 0.01 – 0.05 |
| year | 0.01 *** | 0.01 – 0.02 | 0.01 *** | 0.01 – 0.02 |
| Observations | 263650 | | 251025 | |

*p<0.05   ** p<0.01   *** p<0.001*

**Table S4. Models for children 12-23 m with dietary variables**

| Predictors | Boys 12-24 | | Girls 12-24 | |
|---|---|---|---|---|
| | Estimates | CI | Estimates | CI |
| Increase from Lower Asymptote (a) | 3.98 * | 3.07 – 4.90 | 4.16 * | 3.08 – 5.24 |
| Inflection point (c) | 1.44 * | 0.91 – 1.97 | 1.24 * | 0.70 – 1.77 |
| Lower Asymptote (d) | -3.59 * | -4.21 – -2.97 | -3.55 * | -4.29 – -2.81 |
| Child Age | -0.07 * | -0.08 – -0.05 | -0.07 * | -0.09 – -0.05 |
| Birth Order = 2 | -0.05 | -0.08 – -0.01 | -0.07 * | -0.10 – -0.03 |
| Birth Order > 2 | -0.09 * | -0.14 – -0.04 | -0.10 * | -0.15 – -0.05 |
| Facility Birth | 0.13 * | 0.09 – 0.17 | 0.11 * | 0.08 – 0.15 |
| Facility Birth-Missing | -0.02 | -0.15 – 0.11 | 0.1 | -0.02 – 0.23 |
| > 3 Antenatal Visits | 0.12 * | 0.08 – 0.16 | 0.11 * | 0.08 – 0.15 |
| Antenatal Visits-Missing | 0.11 | 0.01 – 0.21 | 0.16 * | 0.07 – 0.26 |
| > 7 Vaccinations | 0.01 | -0.01 – 0.04 | 0.04 * | 0.02 – 0.07 |
| Number of Children | -0.05 * | -0.07 – -0.04 | -0.05 * | -0.07 – -0.04 |
| Literacy | 0.07 * | 0.03 – 0.11 | 0.04 | 0.00 – 0.08 |
| Primary Education | 0.05 | 0.01 – 0.09 | 0.06 * | 0.02 – 0.10 |
| Secondary Education | 0.21 * | 0.14 – 0.28 | 0.19 * | 0.13 – 0.26 |
| Higher Education | 0.46 * | 0.32 – 0.61 | 0.44 * | 0.29 – 0.58 |
| Mother's Age | 0.23 * | 0.16 – 0.29 | 0.22 * | 0.15 – 0.28 |
| Mother's Age Squared | -0.08 * | -0.11 – -0.05 | -0.03 * | -0.06 – -0.01 |
| Absolute Household Wealth | 0.15 * | 0.11 – 0.18 | 0.13 * | 0.09 – 0.16 |
| Urban | 0.03 | 0.00 – 0.07 | 0.02 | -0.00 – 0.05 |
| Open Defecation-Household | 0 | -0.04 – 0.03 | -0.03 | -0.06 – 0.01 |
| Open Defecation-Cluster | -0.17 | -0.31 – -0.03 | -0.13 | -0.26 – -0.01 |
| Open Defecation > 0.30 | 0.11 | -0.06 – 0.29 | 0.13 | -0.03 – 0.29 |
| Diarrhea-Cluster Level | -0.21 * | -0.32 – -0.10 | -0.14 * | -0.23 – -0.05 |
| Diarrhea-Subdistrict Level | -0.34 | -0.62 – -0.06 | -0.39 * | -0.65 – -0.13 |
| Survey Year | 0.01 * | 0.01 – 0.02 | 0.01 * | 0.01 – 0.02 |
| Ate Dairy in last 24 Hours | 0.12 * | 0.08 – 0.15 | 0.10 * | 0.06 – 0.13 |
| Ate Meat in last 24 Hours | 0.14 * | 0.10 – 0.18 | 0.12 * | 0.08 – 0.16 |
| sub-Saharan Africa d | 0.06 | -0.15 – 0.27 | 0.09 | -0.11 – 0.30 |
| South & Southeast Asia d | -0.55 * | -0.83 – -0.27 | -0.71 * | -0.99 – -0.43 |
| sub-Saharan Africa a | -0.46 * | -0.69 – -0.22 | -0.35 * | -0.59 – -0.11 |
| South & Southeast Asia a | 0.03 | -0.20 – 0.27 | 0.35 | 0.09 – 0.62 |
| Observations | 124676 | | 119890 | |

* p<0.005

**Table S5. Country-level basal HAZestimates**

| | | | | | | | Hindu birth gradient | | | |
| | 24-35 m | | 36-59 m | | 12-23 + Diet | | 24-35 m | | Linear model 24-35 m | |
| CCODE | Boys | Girls | Boys | Girls | Boys | Girls | Boys | Girls | Boys | Girls |
|---|---|---|---|---|---|---|---|---|---|---|
| AL | -3.58 | -5.80 | -2.98 | -3.13 | -3.43 | -3.54 | -3.56 | -5.32 | -2.55 | -2.45 |
| AM | -3.52 | -5.73 | -3.14 | -3.19 | -3.46 | -3.45 | -3.50 | -5.25 | -2.46 | -2.35 |
| AO | -4.07 | -6.25 | NA | NA | -3.74 | -3.75 | -4.06 | -5.81 | -3.17 | -3.00 |
| AZ | -3.90 | -6.08 | -3.57 | -3.55 | -3.56 | -3.69 | -3.88 | -5.61 | -2.90 | -2.76 |
| BD | -4.22 | -6.69 | -3.43 | -3.64 | -4.19 | -4.35 | -4.14 | -6.05 | -3.23 | -3.09 |
| BF | -3.71 | -5.78 | -3.27 | -3.29 | -3.41 | -3.35 | -3.70 | -5.35 | -2.80 | -2.54 |
| BJ | -3.83 | -5.87 | -3.61 | -3.57 | -3.50 | -3.37 | -3.82 | -5.44 | -2.92 | -2.61 |
| BO | -3.84 | -5.97 | -3.15 | -3.19 | -3.78 | -3.69 | -3.83 | -5.51 | -2.82 | -2.61 |
| BR | -3.22 | -5.41 | -2.71 | -2.67 | -3.15 | -3.22 | -3.21 | -4.94 | -2.18 | -2.04 |
| BU | -4.34 | -6.40 | -3.68 | -3.57 | -4.08 | -3.91 | -4.33 | -5.96 | -3.46 | -3.14 |
| CD | -3.83 | -5.82 | -3.73 | -3.63 | -3.44 | -3.30 | -3.82 | -5.39 | -2.93 | -2.56 |
| CF | -3.81 | -5.87 | NA | NA | -3.48 | -3.45 | -3.80 | -5.43 | -2.91 | -2.64 |
| CG | -3.67 | -5.78 | -3.17 | -3.12 | -3.36 | -3.41 | -3.66 | -5.35 | -2.76 | -2.52 |
| CI | -3.55 | -5.72 | -3.25 | -3.30 | -3.40 | -3.19 | -3.55 | -5.28 | -2.64 | -2.47 |
| CM | -3.83 | -5.96 | -3.48 | -3.34 | -3.51 | -3.51 | -3.82 | -5.53 | -2.93 | -2.72 |
| CO | -3.65 | -5.77 | -3.05 | -3.07 | -3.58 | -3.53 | -3.64 | -5.30 | -2.61 | -2.40 |
| DR | -3.35 | -5.47 | -2.71 | -2.66 | -3.27 | -3.19 | -3.33 | -5.00 | -2.30 | -2.09 |
| EG | -3.75 | -5.86 | -3.02 | -3.02 | -3.67 | -3.55 | -3.73 | -5.38 | -2.71 | -2.49 |
| ET | -3.88 | -6.02 | -3.61 | -3.56 | -3.54 | -3.53 | -3.87 | -5.59 | -2.98 | -2.77 |
| GA | -3.76 | -5.85 | -3.35 | -3.19 | -3.51 | -3.50 | -3.75 | -5.42 | -2.86 | -2.60 |
| GH | -3.65 | -5.73 | -3.27 | -3.21 | -3.30 | -3.25 | -3.64 | -5.30 | -2.75 | -2.48 |
| GM | -3.56 | -5.66 | -3.09 | -3.22 | -3.31 | -3.18 | -3.55 | -5.22 | -2.62 | -2.36 |
| GN | -3.51 | -5.62 | -3.36 | -3.32 | -3.08 | -3.06 | -3.50 | -5.19 | -2.59 | -2.36 |
| GU | -4.54 | -6.61 | -3.69 | -3.74 | -4.43 | -4.29 | -4.52 | -6.14 | -3.54 | -3.26 |

| | | | | | | | | | |
|---|---|---|---|---|---|---|---|---|---|
| GY | -3.72 | -5.93 | -2.90 | -2.89 | -3.67 | -3.62 | -3.69 | -5.45 | -2.71 | -2.59 |
| HN | -3.81 | -5.96 | -3.23 | -3.22 | -3.75 | -3.65 | -3.80 | -5.48 | -2.79 | -2.58 |
| HT | -3.45 | -5.55 | -2.78 | -2.70 | -3.28 | -3.18 | -3.44 | -5.08 | -2.43 | -2.17 |
| IA | -4.26 | -6.71 | -3.51 | -3.64 | -4.14 | -4.30 | -4.15 | -6.01 | -3.28 | -3.12 |
| JO | -3.53 | -5.67 | -2.96 | -3.03 | -3.41 | -3.37 | -3.51 | -5.19 | -2.47 | -2.29 |
| KE | -3.82 | -5.85 | -3.34 | -3.21 | -3.53 | -3.47 | -3.81 | -5.42 | -2.92 | -2.60 |
| KH | -3.98 | -6.43 | -3.48 | -3.58 | -3.99 | -4.17 | -3.91 | -5.81 | -2.98 | -2.82 |
| KK | -3.66 | -5.83 | -3.18 | -3.14 | -3.52 | -3.47 | -3.64 | -5.35 | -2.62 | -2.47 |
| KM | -3.84 | -5.85 | -3.16 | -3.14 | -3.69 | -3.43 | -3.83 | -5.41 | -2.94 | -2.60 |
| KY | -3.96 | -6.10 | -3.39 | -3.43 | -3.78 | -3.66 | -3.94 | -5.63 | -2.93 | -2.73 |
| LB | -3.63 | -5.58 | -3.35 | -3.15 | -3.28 | -3.09 | -3.63 | -5.15 | -2.72 | -2.31 |
| LS | -4.03 | -6.12 | -3.71 | -3.58 | -3.69 | -3.63 | -4.02 | -5.69 | -3.14 | -2.87 |
| MA | -3.35 | -5.57 | -2.94 | -2.86 | -3.15 | -3.18 | -3.34 | -5.10 | -2.32 | -2.21 |
| MB | -3.21 | -5.31 | -2.76 | -2.79 | -3.13 | -3.04 | -3.19 | -4.84 | -2.14 | -1.91 |
| MD | -4.17 | -6.21 | -3.89 | -3.72 | -3.85 | -3.72 | -4.17 | -5.78 | -3.28 | -2.97 |
| ML | -3.81 | -5.88 | -3.50 | -3.52 | -3.43 | -3.40 | -3.79 | -5.44 | -2.90 | -2.63 |
| MM | -4.18 | -6.58 | -3.49 | -3.64 | -4.00 | -4.06 | -4.10 | -5.93 | -3.15 | -2.90 |
| MR | -3.57 | -5.57 | -3.28 | -3.15 | -3.36 | -3.29 | -3.56 | -5.14 | -2.64 | -2.31 |
| MV | -4.04 | -6.56 | -3.13 | -3.33 | -4.10 | -4.30 | -3.95 | -5.91 | -3.06 | -2.95 |
| MW | -4.04 | -6.07 | -3.76 | -3.64 | -3.77 | -3.65 | -4.03 | -5.64 | -3.15 | -2.83 |
| MZ | -3.98 | -6.03 | -3.55 | -3.38 | -3.67 | -3.60 | -3.97 | -5.59 | -3.08 | -2.78 |
| NC | -3.55 | -5.71 | -3.04 | -3.06 | -3.46 | -3.44 | -3.53 | -5.24 | -2.53 | -2.36 |
| NG | -4.06 | -6.08 | -3.77 | -3.60 | -3.68 | -3.60 | -4.05 | -5.65 | -3.16 | -2.84 |
| NI | -3.91 | -5.99 | -3.68 | -3.40 | -3.64 | -3.65 | -3.91 | -5.56 | -3.01 | -2.76 |
| NM | -3.78 | -5.86 | -3.42 | -3.24 | -3.55 | -3.48 | -3.78 | -5.43 | -2.88 | -2.62 |
| NP | -4.09 | -6.57 | -3.42 | -3.61 | -3.94 | -4.15 | -3.97 | -5.86 | -3.10 | -2.98 |
| PE | -4.05 | -6.16 | -3.39 | -3.40 | -4.01 | -3.94 | -4.03 | -5.69 | -3.00 | -2.78 |
| PK | -4.30 | -6.80 | -3.76 | -3.79 | -4.15 | -4.27 | -4.24 | -6.16 | -3.31 | -3.18 |

| | | | | | | | | | |
|----|------|------|------|------|------|------|------|------|------|------|
| PY | -3.52 | -5.65 | -2.76 | -2.74 | NA | NA | -3.50 | -5.17 | -2.50 | -2.30 |
| RW | -4.10 | -6.15 | -3.72 | -3.57 | -3.86 | -3.75 | -4.08 | -5.71 | -3.20 | -2.91 |
| SL | -3.60 | -5.70 | -3.37 | -3.23 | -3.34 | -3.18 | -3.59 | -5.26 | -2.68 | -2.42 |
| SN | -3.42 | -5.57 | -3.05 | -2.96 | -3.15 | -3.18 | -3.41 | -5.14 | -2.51 | -2.30 |
| ST | -3.51 | -5.51 | -3.06 | -3.02 | -3.37 | -3.23 | -3.51 | -5.08 | -2.57 | -2.21 |
| SZ | -3.93 | -6.03 | -3.45 | -3.34 | -3.67 | -3.59 | -3.92 | -5.60 | -3.03 | -2.80 |
| TD | -3.94 | -6.06 | -3.56 | -3.58 | -3.48 | -3.52 | -3.93 | -5.64 | -3.03 | -2.82 |
| TG | -3.48 | -5.52 | -3.26 | -3.19 | -3.19 | -3.08 | -3.48 | -5.09 | -2.56 | -2.26 |
| TJ | -3.82 | -6.03 | -3.31 | -3.36 | -3.57 | -3.58 | -3.80 | -5.56 | -2.77 | -2.64 |
| TL | -4.55 | -6.80 | -3.83 | -3.88 | -4.54 | -4.40 | -4.48 | -6.16 | -3.55 | -3.16 |
| TR | -3.40 | -5.56 | -3.05 | -3.17 | NA | NA | -3.38 | -5.08 | -2.36 | -2.20 |
| TZ | -4.10 | -6.19 | -3.81 | -3.68 | -3.75 | -3.71 | -4.09 | -5.75 | -3.21 | -2.95 |
| UG | -3.81 | -5.84 | -3.42 | -3.17 | -3.57 | -3.43 | -3.80 | -5.41 | -2.91 | -2.59 |
| UZ | -4.03 | -6.05 | NA | NA | -3.87 | -3.81 | -4.01 | -5.57 | -3.05 | -2.71 |
| YE | -3.98 | -6.36 | -3.60 | -3.69 | -3.94 | -3.90 | -3.97 | -5.90 | -2.97 | -3.02 |
| ZA | -3.95 | -6.20 | -3.58 | -3.53 | -3.61 | -3.63 | -3.94 | -5.77 | -3.07 | -3.00 |
| ZM | -4.16 | -6.20 | -3.78 | -3.59 | -3.88 | -3.74 | -4.16 | -5.76 | -3.27 | -2.96 |
| ZW | -3.77 | -5.84 | -3.27 | -3.12 | -3.45 | -3.39 | -3.76 | -5.40 | -2.88 | -2.59 |

**Figure S1. Distribution of 225 DHS surveys by proportion of non-missing values with absolute magnitudes greater than 6.**

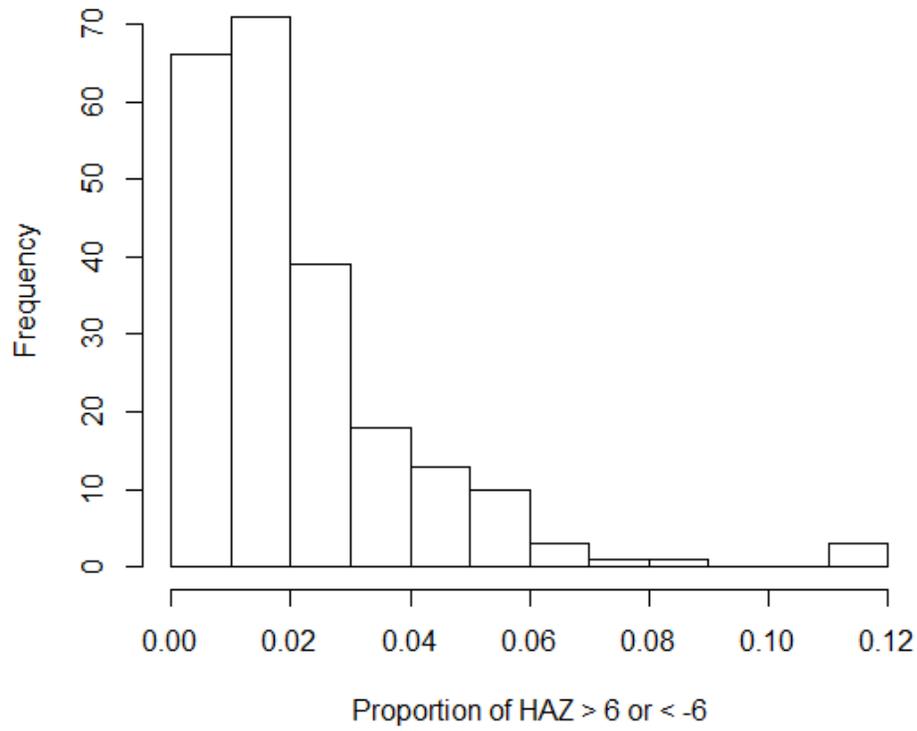

**Figure S2.  Country-level bHAZ estimates compared to Survey-level bHAZ estimates.**

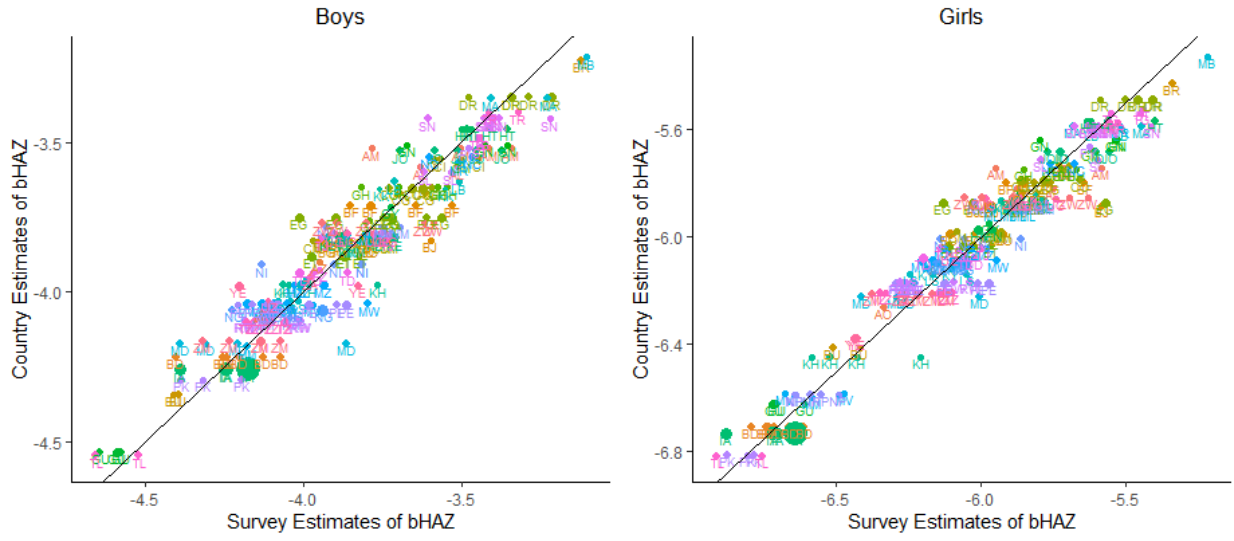